\documentclass[12pt]{article}
\usepackage{epsfig}
\def\vev#1{\left\langle #1 \right\rangle}

\newfont{\Bbb}{msbm10 scaled 1200}     %instead of eusb10
\newcommand{\mathbb}[1]{\mbox{\Bbb #1}}

\def\lbldef#1#2{\expandafter\gdef\csname #1\endcsname {#2}}

\def\href#1#2{#2}

\newcommand{\beq}{\begin{equation}}
\newcommand{\eeq}{\end{equation}}
\newcommand{\ber}{\begin{eqnarray}}
\newcommand{\eer}{\end{eqnarray}}

\newcommand{\beqar}{\begin{eqnarray}}

\newcommand{\eeqar}{\end{eqnarray}}
\begin{document}
\baselineskip=15.5pt
\renewcommand{\theequation}{\arabic{section}.\arabic{equation}}
\pagestyle{plain}
\setcounter{page}{1}
%\renewcommand{\thefootnote}{\fnsymbol{footnote}}
%--------+---------+---------+---------+---------+---------+---------+
%Title page
\begin{titlepage}

\leftline{\tt hep-th/0301133}

\vskip -.5cm

\rightline{\small{\tt CALT-68-2420}}

\begin{center}

\vskip 2 cm

{\Large {Quantum aspects of Seiberg-Witten map}}

\vskip .5cm

{\Large{in noncommutative Chern-Simons theory}}

\vskip 1.5cm
{\large Kirk Kaminsky\footnote{
e-mail address:\ \ kaminsky@theory.caltech.edu}, \
Yuji Okawa\footnote{
e-mail address:\ \ okawa@theory.caltech.edu} \
and \ Hirosi Ooguri\footnote{
e-mail address:\ \ ooguri@theory.caltech.edu}}

\vskip 1.0cm

{{\it California Institute of Technology}}

\smallskip

{{\it Mail Code 452-48}}

\smallskip

{{\it Pasadena, CA 91125 USA}}

\vskip 2cm

{\bf Abstract}
\end{center}

\noindent
Noncommutative Chern-Simons theory can be classically mapped
to commutative Chern-Simons theory by the Seiberg-Witten map.
We provide evidence that the equivalence persists at the quantum
level by computing two and three-point functions of field strengths
on the commutative side and their Seiberg-Witten transforms
on the noncommutative side to the first nontrivial order in
perturbation theory.

\end{titlepage}

\newpage

%--------+---------+---------+---------+---------+---------+---------+
%Body

%%%%% Section 1 %%%%%
\section{Introduction}
\setcounter{equation}{0}

The Seiberg-Witten limit \cite{Seiberg:1999vs}
is an interesting limit of open string theory
with a constant NS-NS $B$ field, in which
open string dynamics reduces to a gauge theory defined
on a noncommutative space. The theory in the limit
can also be described in terms of fields defined
on a commutative space. It was shown in \cite{Seiberg:1999vs}
that these two descriptions are related to each other
by a field redefinition called the Seiberg-Witten
map. When the gauge group is $U(1)$, the Seiberg-Witten
map has been obtained explicitly
in \cite{Okawa:2001mv, Mukhi:2001vx, Liu:2001pk}
by studying
the coupling of D-branes to the Ramond-Ramond potentials
and by evaluating it in the Seiberg-Witten limit.
It was shown that the field strength on the commutative
space is expressed in terms of the open Wilson line
\cite{Ishibashi:1999hs, Das:2000md, Gross:2000ba}
with appropriate insertions of operators on the
noncommutative space.

Although the two descriptions are equivalent,
fields on a noncommutative space are often more
convenient in studying theories in the
Seiberg-Witten limit. Actions expressed
in terms of fields on a commutative space typically
become nonpolynomial in the limit \cite{Seiberg:1999vs}
and their closed forms are not known in general, though
some constraints on possible terms in such actions
have been studied in cases when they are realized as
limits of string theory
\cite{Okawa:1999cm, Okawa:2000em, Das:2001xy}.
The lack of our
understanding of actions on commutative spaces has prevented us from
studying whether the equivalence implied by the Seiberg-Witten
map holds at the quantum level.

One interesting case in which gauge theory
actions are known in both descriptions is
Chern-Simons theory in three dimensions. Its
action in the noncommutative space is given by
\begin{equation}
  S_{NCCS} = \frac{1}{2} \int d^3 x~ \epsilon^{\mu \rho \nu}
  {\rm tr} \left[
    A_\mu \ast \partial_\rho A_\nu
    - \frac{2ig}{3} A_\mu \ast A_\rho \ast A_\nu
  \right],
\end{equation}
where the product is the standard star-product:
\begin{equation}
  f(x) \ast g(x) = \exp \left( \frac{i \theta^{\mu \nu}}{2}
  \frac{\partial}{\partial \xi^\mu}
  \frac{\partial}{\partial \zeta^\nu} \right)
  f(x+\xi) g(x+\zeta) \Bigg|_{\xi=\zeta=0},
\end{equation}
and ${\rm tr}$ is over the gauge group indices.
It was shown \cite{Grandi:2000av, Polychronakos:2002pm}
that this action, when expressed
in terms of gauge field $a_\mu(x)$ on a commutative
space via the Seiberg-Witten map, becomes the standard
Chern-Simons action. The proof of this statement is
based on the differential equation characterizing the
Seiberg-Witten map and holds for any $U(N)$ gauge group.

This is an interesting case for various reasons.
First of all, since the actions are known in both
descriptions
and they both appear renormalizable, we can
discuss the question of whether the equivalence
of the two descriptions at the classical level
can be extended to the quantum level.
In this regard, there is an interesting puzzle.
When the gauge group is $U(1)$, the Chern-Simons
theory on the commutative space is trivial
while its noncommutative counterpart has a cubic
interaction. The latter theory seems to depend
nontrivially on the coupling constant $g$, while
the corresponding parameter for the former can be
rescaled away. This casts some doubt on the quantum
equivalence of the two. One of the motivations
of this paper is to understand whether the equivalence
in fact breaks down at the quantum level. We will
compute correlation functions of open Wilson lines
on the noncommutative space to the first nontrivial
order in perturbation theory\footnote
{Perturbative aspects of noncommutative Chern-Simons theory
have been studied in \cite{Bichl:2000bq, Chen:2000ak}.}
and find that the equivalence persists
at the quantum level.

The $U(1)$ Chern-Simons theory on the noncommutative
space is expected to describe aspects
of fractional quantum Hall fluid
\cite{Susskind:2001fb, Polychronakos:2001mi, Hellerman:2001rj},
and correlation functions of the open Wilson lines
we will discuss in this paper play important roles
in this context.
Moreover it is known that such a theory is realized
in a certain configuration of D-branes in string theory
\cite{Bergman:2001qg, Hellerman:2001yv}.
We hope that the results in this paper
shed some light on these issues.

When the gauge group is $U(N)$,
an explicit form of the Seiberg-Witten map
has not been derived
in the sense of the works
\cite{Okawa:2001mv, Mukhi:2001vx, Liu:2001pk}.
However, the map between a certain subset of observables
on commutative and noncommutative sides
can be extended to the $U(N)$ case.
The generalization of our computations
is straightforward,
and we find that the equivalence holds
for the $U(N)$ case as well.

The organization of the paper is as follows.
In Section 2 we review the derivation of an exact expression
for the Seiberg-Witten map and introduce a regularization
for the composite operators appearing in the expression.
We then calculate two and three-point functions
of field strengths on the commutative side
and their Seiberg-Witten transforms on the noncommutative side
in perturbation theory in Section 3.
We present our conclusions in Section 4,
wherein we also discuss our generalization to the $U(N)$ case.
Our conventions and Feynman rules are summarized in
Appendix A, and some details of the computations in Section 3
are given in Appendix B.

%%%%% Section 2 %%%%%
\section{Seiberg-Witten map and its regularization}
\setcounter{equation}{0}

%%%%% Subsection 2.1 %%%%%
\subsection{Exact expression for the Seiberg-Witten map}
\setcounter{equation}{0}

An exact expression for the Seiberg-Witten map
in arbitrary dimensions
was derived by studying the Ramond-Ramond couplings
of noncommutative gauge theory
\cite{Okawa:2001mv, Mukhi:2001vx, Liu:2001pk}.\footnote{
The Seiberg-Witten map from noncommuting to commuting variables
is related to the Lagrange to Euler map in fluid mechanics
\cite{Jackiw:2002pn, Susskind:2001fb}.}
It takes a simple form in three dimensions.
If we choose a coordinate system such that
only $\theta^{12}$ and $\theta^{21}$ are nonvanishing,
the Seiberg-Witten map is given by
\begin{eqnarray}
  f_{12}(k)
  &=& -\frac{1}{g \theta^{12}}
    \left[ W(k) - (2\pi)^3 \delta^{(3)}(k) \right],
\nonumber \\
  f_{0i}(k) &=& O_{0i}(k) \quad {\rm for} \quad i=1,2,
\label{SW-map}
\end{eqnarray}
where $f_{\mu \nu}(k)$ is the field strength
on the commutative side in momentum space,
$W(k)$ is an open Wilson line
\begin{equation}
  W(k) = \int d^3 x~ P_* \exp \left[ ig \int_0^1 d \sigma~
         l^\mu A_\mu (x + l \sigma) \right] \ast e^{ikx}
\end{equation}
with
\begin{equation}
  l^\mu \equiv (k \theta)^\mu = k_\nu \theta^{\nu \mu},
\label{l-definition}
\end{equation}
and $O_{\mu \nu}(k)$ is defined by
\begin{equation}
  O_{\mu \nu}(k) = \int d^3 x~ P_* \exp \left[ ig \int_0^1 d \sigma~
    l^\mu A_\mu (x + l \sigma) \right]
    \ast F_{\mu \nu} (x) \ast e^{ikx}
\label{O-definition}
\end{equation}
with
\begin{equation}
  F_{\mu \nu} = \partial_\mu A_\nu - \partial_\nu A_\mu
                -ig A_\mu \ast A_\nu +ig A_\nu \ast A_\mu.
\end{equation}
Our convention for the path-ordered exponential is as follows:
\begin{eqnarray}
  && P_* \exp \left[ ig \int_0^1 d \sigma~
         l^\mu A_\mu (x + l \sigma) \right]
\nonumber \\
  && = 1 + ig \int_0^1 d \sigma_1~ l \cdot A (x + l \sigma_1)
\nonumber \\ && \qquad
    {}+ (ig)^2 \int_0^1 d \sigma_1 \int_0^{\sigma_1} d \sigma_2~
    l \cdot A (x + l \sigma_1) \ast l \cdot A (x + l \sigma_2)
    + O(g^3).
\end{eqnarray}
The Seiberg-Witten map can also be written in the covariant form
$f_{\mu \nu}(k) = O_{\mu \nu}(k)$ for $\mu, \nu = 0,1,2$,
as originally conjectured in \cite{Liu:2000mj}.
However, the expression (\ref{SW-map}) is more convenient
for our perturbative computations.

In \cite{Okawa:2001mv}
an expression of $f_{\mu \nu}(k; A_\mu, \theta)$
for arbitrary dimensions was constructed which
\noindent
($a$) is gauge invariant,
\begin{equation}
  f_{\mu \nu}(k; A_\mu+\partial_\mu \lambda -
   ig A_\mu \ast \lambda + ig \lambda \ast A_\mu, \theta)
  = f_{\mu \nu}(k; A_\mu, \theta),
\end{equation}

\noindent
($b$) obeys the Bianchi identity for the ordinary gauge theory:
\begin{equation}
  k_\mu f_{\rho \nu}(k) + k_\rho f_{\nu \mu}(k)
  + k_\nu f_{\mu \rho}(k) = 0,
\end{equation}

\noindent
($c$) and satisfies the initial condition,
\begin{equation}
  \lim_{\theta \to 0} f_{\mu \nu}(k; A_\mu, \theta)
  = \int d^3 x~ \left[ \partial_\mu A_\nu (x)
                     - \partial_\nu A_\mu (x) \right] e^{ikx}.
\label{initial-condition}
\end{equation}
A proof for arbitrary dimensions
was given in \cite{Okawa:2001mv},
but it is much easier to see that
$f_{\mu \nu}(k)$ in three dimensions
defined by (\ref{SW-map}) satisfies these three conditions.
The gauge invariance is guaranteed by the relation (\ref{l-definition})
\cite{Ishibashi:1999hs}, and the initial condition is easily verified.
It is instructive to verify that
the Seiberg-Witten map (\ref{SW-map}) satisfies
the Bianchi identity
$k_0 f_{12}(k) + k_1 f_{20}(k) + k_2 f_{01}(k) = 0$.
Since $k_0 \delta^{(3)} (k) = 0$,
what we need to show is
$ k_0 W(k) + g (k \theta)^\mu O_{0 \mu}(k) = 0$.
This can be shown as follows:
\begin{eqnarray}
  && k_0 W(k)
  =  -i \int d^3 x~ P_* \exp \left[ ig \int_0^1 d \sigma~
     l \cdot A (x + l \sigma) \right] \ast \partial_0 (e^{ikx})
\nonumber \\
  && =  -g \int d^3 x~ P_* \Biggl[ \exp \left[ ig \int_0^1 d \sigma~
     l \cdot A (x + l \sigma) \right]
     \int_0^1 d \sigma'~ l^\mu \partial_0 A_\mu (x + l \sigma') \Biggr]
     \ast e^{ikx}
\nonumber \\
  && =  -g \int d^3 x~ P_* \Biggl[ \exp \left[ ig \int_0^1 d \sigma~
     l \cdot A (x + l \sigma) \right]
\nonumber \\ && \qquad \qquad \times
     \int_0^1 d \sigma'~
     l^\mu \{ F_{0 \mu} (x + l \sigma') + D_\mu A_0 (x + l \sigma') \}
     \Biggr] \ast e^{ikx}
\nonumber \\
  && =  -g \int d^3 x~ P_* \exp \left[ ig \int_0^1 d \sigma~
     l \cdot A (x + l \sigma) \right]
     \ast l^\mu F_{0 \mu} (x) \ast e^{ikx}
\nonumber \\
  && = -g (k \theta)^\mu O_{0 \mu}(k),
\end{eqnarray}
where $D_\mu A_0
= \partial_\mu A_0 -ig A_\mu \ast A_0 + ig A_0 \ast A_\mu$.
We integrated by parts in the first step,
and then used the following identities:
\begin{eqnarray}
     \int d^3 x~ P_* \Biggl[ \exp \left[ ig \int_0^1 d \sigma~
     l \cdot A (x + l \sigma) \right]
     \int_0^1 d \sigma'~
     l^\mu  D_\mu A_0 (x + l \sigma') \}
     \Biggr] \ast e^{ikx} = 0,
\end{eqnarray}
which was shown in the appendix of \cite{Okawa:2000sh}
for the conservation of the energy-momentum tensor
derived in the paper, and
\begin{eqnarray}
     && \int d^3 x~ P_* \Biggl[ \exp \left[ ig \int_0^1 d \sigma~
     l \cdot A (x + l \sigma) \right]
     {\cal O} (x + l \sigma') \Biggr] \ast e^{ikx}
\nonumber \\
     &=& \int d^3 x~ P_* \exp \left[ ig \int_0^1 d \sigma~
     l \cdot A (x + l \sigma) \right]
     \ast {\cal O} (x) \ast e^{ikx}
\end{eqnarray}
for $0 \le \sigma' \le 1$,
which is one of the basic properties of a straight open Wilson line
\cite{Gross:2000ba}.

It is well-known that the Seiberg-Witten map is not unique.
However, the ambiguity pointed out in \cite{Asakawa:1999cu} is absent
when the dimension of noncommutative directions is two,
and the definition of the noncommutative gauge field is essentially
unique in the Seiberg-Witten limit.
The definition of the commutative gauge field may in general
admit some ambiguity even in the Seiberg-Witten limit,
but we assume that the expression (\ref{SW-map})
provides the map between noncommutative Chern-Simons theory
and commutative Chern-Simons theory.

%%%%% Subsection 2.2 %%%%%
\subsection{Regularization}

When we compute correlation functions
of $W(k)$ and $O_{\mu \nu}(k)$,
we need to regularize these composite operators.
A pure open Wilson line $W(k)$ is expanded in $g$ as follows:
\begin{eqnarray}
  W(k) &=& \int d^3 x~ e^{ikx}
  + ig \int d^3 x \int_0^1 d \sigma~
    l \cdot A (x + l \sigma) \ast e^{ikx}
\nonumber \\
  && {}+ (ig)^2 \int d^3 x
    \int_0^1 d \sigma_1 \int_0^{\sigma_1} d \sigma_2~
    l \cdot A (x + l \sigma_1) \ast l \cdot A (x + l \sigma_2)
    \ast e^{ikx}
  + O(g^3)
\nonumber \\
  &=& (2\pi)^3 \delta^{(3)}(k) + ig~ l \cdot A(k)
\nonumber \\
  &&  {}+ \frac{(ig)^2}{2} \int d^3 x \int_0^1 d \sigma~
    l \cdot A (x + l \sigma) \ast l \cdot A(x) \ast e^{ikx}
    + O(g^3).
\end{eqnarray}
The expression of $W(k)$ up to this order is sufficient
for our perturbative computations in the next section.
We regularize the composite operator at $O(g^2)$ as follows:
\begin{equation}
  \frac{(ig)^2}{2} \int d^3 x
  \int_\epsilon^{1-\epsilon} d \sigma~
  l \cdot A (x + l \sigma) \ast l \cdot A(x) \ast e^{ikx}.
\end{equation}
Note that in addition to the expected singularity
arising when $\sigma \to 0$,
a singularity also arises when $\sigma \to 1$ since
\begin{eqnarray}
  && \int d^3 x~
  l \cdot A (x + l \sigma) \ast l \cdot A(x) \ast e^{ikx}
  = \int d^3 x~
  l \cdot A(x) \ast e^{ikx} \ast l \cdot A (x + l \sigma)
\nonumber \\
  && = \int d^3 x~
  l \cdot A(x) \ast l \cdot A (x + l \sigma-l) \ast e^{ikx},
\end{eqnarray}
where we have used the basic identities
\begin{equation}
  \int d^3 x~ f(x) \ast g(x) = \int d^3 x~ g(x) \ast f(x),
\label{cyclic}
\end{equation}
for any functions $f(x)$ and $g(x)$ which decay at infinity, and
\begin{equation}
  e^{ikx} \ast f(x+l) = f(x) \ast e^{ikx},
\label{shift}
\end{equation}
for any $C^{\infty}$ function $f(x)$.
This regularization is natural for the following reason.
In \cite{Okawa:2000sh}
it was shown how a straight open Wilson line arises from
the computation of disk amplitudes in string theory.
The integral over $\sigma$ comes from the integral over
a position of an open string vertex operator along the boundary.
Since point-splitting regularization
on the world-sheet boundary
produces noncommutative gauge theory in space-time
\cite{Seiberg:1999vs},
it is natural to use point-splitting regularization
for the integral over $\sigma$ as well.

The operator $O_{\mu \nu}(k)$ (\ref{O-definition})
can also be expanded in $g$ as follows:
\begin{eqnarray}
  && O_{\mu \nu}(k)
  = \int d^3 x~
  \left[ \partial_\mu A_\nu (x) - \partial_\nu A_\mu (x) \right]
  \ast e^{ikx}
\nonumber \\
  && \qquad -ig \int d^3 x~
  \left[ A_\mu (x) \ast A_\nu (x) - A_\nu (x) \ast A_\mu (x) \right]
  \ast e^{ikx}
\nonumber \\
  && \qquad +ig \int d^3 x \int_0^1 d \sigma~
  l \cdot A (x + l \sigma) \ast
  \left[ \partial_\mu A_\nu (x) - \partial_\nu A_\mu (x) \right]
  \ast e^{ikx}
  + O(g^2).
\nonumber \\
\label{expanded-O}
\end{eqnarray}
The integral over $\sigma$ in the last line can be regularized
by taking the integration range from $\epsilon$ to $1-\epsilon$
as before.
The commutator term in the second line is regularized as
\begin{equation}
  -ig \int d^3 x~
  \left[ A_\mu (x + l\epsilon) \ast A_\nu (x)
  - A_\nu (x + l \epsilon) \ast A_\mu (x) \right] \ast e^{ikx}.
\label{commutator-regularization}
\end{equation}
Note that only the difference in two arguments matters.
For example,
\begin{equation}
  \int d^3 x~
  A_\mu (x) \ast A_\nu (x - l \epsilon) \ast e^{ikx}
  = \int d^3 x'~
  A_\mu (x' + l \epsilon) \ast A_\nu (x') \ast e^{ikx'}
\end{equation}
with $x'=x - l \epsilon$ because of the fact that
$l \cdot k = k_\mu \theta^{\mu \nu} k_\nu =0$.

This regularization for the commutator term
is natural for the following reason.
As is well-known,
the commutator terms in the field strength arise from
surface terms of the path-ordered integrals over positions
of open string vertex operators
\cite{Seiberg:1999vs, Okawa:2000sh}.
Therefore, if we use point-splitting regularization
for the integral over $\sigma$, commutator terms should also
be regularized correspondingly.
The relation between the commutator term and the surface term
in (\ref{expanded-O}) at $O(g)$ can be seen,
for example, by looking at the Bianchi identity
$ k_0 W(k) + g l^\mu O_{0 \mu}(k) = 0$.
Since $W(k)$ does not depend on $A_0(x)$,
the part of $l^\mu O_{\mu 0} (k)$ which depends on $A_0 (x)$
must cancel by itself. This can be seen for the regularized
$O_{\mu \nu} (k)$ at order $g$ as follows:
\begin{eqnarray}
  && l^\mu \Biggl[
  ig \int d^3 x \int_\epsilon^{1-\epsilon} d \sigma~
  l \cdot A (x + l \sigma) \ast \partial_\mu A_0 (x)
  \ast e^{ikx} \Biggr]
\nonumber \\
  && = ig \int d^3 x \int_\epsilon^{1-\epsilon} d \sigma~
  l \cdot \partial A_0 (x) \ast l \cdot A (x + l \sigma -l)
  \ast e^{ikx}
\nonumber \\
  && = ig \int d^3 x \int_\epsilon^{1-\epsilon} d \sigma~
  l \cdot \partial A_0 (x + l - l \sigma) \ast l \cdot A (x)
  \ast e^{ikx}
\nonumber \\
  && = -ig \int d^3 x \int_\epsilon^{1-\epsilon} d \sigma~
  \frac{\partial}{\partial \sigma} A_0 (x + l - l \sigma)
  \ast l \cdot A (x) \ast e^{ikx}
\nonumber \\
  && = -ig \int d^3 x
  \left\{ A_0 (x + l \epsilon) - A_0 (x + l - l \epsilon) \right\}
  \ast l \cdot A (x) \ast e^{ikx}
\nonumber \\
  && = -ig \int d^3 x
  \left\{ A_0 (x + l \epsilon) \ast l \cdot A (x)
         - l \cdot A (x) \ast A_0 (x - l \epsilon) \right\}
  \ast e^{ikx}
\nonumber \\
  && = l^\mu \Biggl[
  -ig \int d^3 x
  \left\{ A_0 (x + l \epsilon) \ast A_\mu (x)
         - A_\mu (x + l \epsilon) \ast A_0 (x) \right\}
  \ast e^{ikx} \Biggr],
\nonumber \\
\end{eqnarray}
where we have used the identities (\ref{cyclic}) and (\ref{shift}),
and the change of variables
$x' = x + l \sigma -l$.
It is not difficult to see that
the remaining part of the Bianchi identity
which is independent of $A_0 (x)$ also holds
for a finite $\epsilon$ up to the current order in $g$.
We therefore conclude that our regularization
for the commutator (\ref{commutator-regularization})
is in accord with
point-splitting regularization of the integral over $\sigma$.

To summarize, we use the following regularized operators
in terms of the gauge field in momentum space
to compute correlation functions:
\begin{eqnarray}
  W(k) &=& (2\pi)^3 \delta^{(3)}(k) + ig~ l \cdot A(k)
\nonumber \\
  &&  + \frac{(ig)^2}{2} \int_\epsilon^{1-\epsilon} d \sigma
        \int \frac{d^3 p}{(2 \pi)^3}~
        e^{-i k \times p \sigma + \frac{i}{2} k \times p}~
        l \cdot A (p)~ l \cdot A(k-p)
    + O(g^3),
\nonumber \\
  O_{\mu \nu}(k)
  &=& -i k_\mu A_\nu (k) +i k_\nu A_\mu (k)
\nonumber \\
  && -ig \int \frac{d^3 p}{(2\pi)^3}~
     e^{-i k \times p \epsilon + \frac{i}{2} k \times p}~
  \left[ A_\mu (p) A_\nu (k-p) - A_\nu (p) A_\mu (k-p) \right]
\nonumber \\
  && +ig \int_\epsilon^{1-\epsilon} d \sigma
     \int \frac{d^3 p}{(2\pi)^3}~
     e^{-i k \times p \sigma + \frac{i}{2} k \times p}~
\nonumber \\
  && \qquad \qquad  {}\times l \cdot A (p)
     \left[ -i (k-p)_\mu A_\nu (k-p) +i (k-p)_\nu A_\mu (k-p) \right]
\nonumber \\
  && {}+ O(g^2),
\end{eqnarray}
where we have introduced the notation
$k \times p \equiv k_\mu \theta^{\mu \nu} p_\nu$.

%%%%% Section 3 %%%%%
\section{Computations of correlation functions}
\setcounter{equation}{0}

%%%%% Subsection 3.1 %%%%%
\subsection{Correlation functions on the commutative side}

Correlation functions of field strengths can be easily calculated
on the commutative side, where the action is given by
\begin{equation}
  S_{CS} = \frac{1}{2} \int d^3 x~ \epsilon^{\mu \rho\nu}
           a_\mu \partial_\rho a_\nu.
\end{equation}
Since the field strength $f_{\mu \nu}$ can be expressed as
\begin{equation}
  f_{\mu \nu} (x) = \epsilon_{\mu \nu \rho}
                    \frac{\delta S_{CS}}{\delta a_\rho (x)},
\end{equation}
correlation functions can be easily evaluated
by using the Schwinger-Dyson equations.

Correlation functions containing only $f_{12}$ vanish because
\begin{eqnarray}
  && \vev{f_{12}(x_1) f_{12}(x_2) \ldots f_{12}(x_n)}
  = \int {\cal D} a~ f_{12}(x_1) f_{12}(x_2) \ldots f_{12}(x_n)~
       e^{i S_{CS}}
\nonumber \\
  && = -i \int {\cal D} a~ f_{12}(x_2) \ldots f_{12}(x_n)~
       \frac{\delta}{\delta a_0 (x_1)} e^{i S_{CS}}
\nonumber \\
  && = -i \int {\cal D} a~ \frac{\delta}{\delta a_0 (x_1)} \left[
       f_{12}(x_2) \ldots f_{12}(x_n)~ e^{i S_{CS}} \right]
  = 0.
\end{eqnarray}
The two-point function of $f_{12}$ and $f_{0i}$ is nonvanishing
and is given by
\begin{eqnarray}
  && \vev{f_{12} (x) f_{0i} (y)}
  = \int {\cal D} a~ f_{12}(x) f_{0i}(y)~ e^{i S_{CS}}
\nonumber \\
  && = -i \int {\cal D} a~ f_{0i}(y)~
       \frac{\delta}{\delta a_0 (x)} e^{i S_{CS}}
  = i \int {\cal D} a~
       \frac{\delta f_{0i}(y)}{\delta a_0 (x)} e^{i S_{CS}}
  = i \frac{\partial}{\partial x^i} \delta^{(3)} (x-y)
\end{eqnarray}
for $i=1,2$.
In momentum space, it is given by
\begin{equation}
  \vev{ f_{12}(k) f_{0i}(k') }
  = (2\pi)^3 k_i \delta^{(3)}(k+k') \quad {\rm for} \quad i=1,2.
\label{commutative-two-point}
\end{equation}
We will calculate the corresponding gauge-invariant observables
on the noncommutative side to see if these results
are reproduced.

%%%%% Subsection 3.2 %%%%%
\subsection{$\vev{W(k)W(k')}$}

One-point functions of $W(k)$ or $O_{\mu \nu}(k)$
are rather trivial because the length of an open Wilson line
is proportional to the momentum $k$,
while momentum conservation enforces $k=0$.
We cannot completely exclude possible subtleties
arising from possible short-distance singularities,
but in this paper we would rather study
two-point and three-point functions
which are more interesting.

Let us begin with the two-point function $\vev{W(k)W(k')}$.
Since $k'=-k$ from momentum conservation,
all of the gauge fields $A_\mu$ in each of two open Wilson lines
are contracted with the same vector $(k \theta)^\mu$ up to a sign.
If we choose a coordinate system such that $k_2=0$
by a rotation in the $(x^1,x^2)$-plane,
only $A_2$ appears in the open Wilson lines
since the only nonvanishing component
of $(k \theta)^\mu$ is $\mu=2$.
Therefore, the two-point function $\vev{W(k)W(k')}$
consists of correlation functions involving only $A_2$,
$\vev{A_2(p_1) A_2(p_2) \ldots A_2(p_n)}$.

In noncommutative Chern-Simons theory, the propagator
$\vev{A_2(p) A_2(q)}$ vanishes in the Landau gauge.\footnote
{Feynman rules are summarized in Appendix A.}
Therefore, we cannot contract any pair of gauge fields
coming from the two Wilson lines directly.
Contractions such as the ones shown in Figure 1 are prohibited.
%%%%% Figure 1 %%%%%
\begin{figure}
\centerline{\epsfxsize=2in\epsfbox{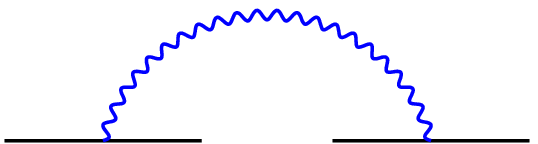} \hspace{1cm}
\epsfxsize=2in\epsfbox{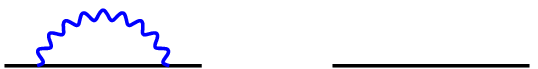}}
\caption{Vanishing contractions.}
\end{figure}
%%%%%%%%%%%%%%%%%%%%
This rule also applies to $\vev{W(k)O_{\mu \nu}(k')}$
which we will discuss in the next subsection. From
this it immediately follows that $O(g^2)$ contribution
to $\vev{W(k)W(k')}$ vanishes:
\begin{equation}
  \vev{W(k)W(k')}
  = (2\pi)^6 \delta^{(3)}(k) \delta^{(3)}(k') + O(g^4).
\end{equation}

Furthermore, even if the cubic vertices are used,
we end up with at least one contraction of $\vev{A_2(p) A_2(q)}$
unless there is at least one internal loop.
Let us take the diagram shown in Figure 2 as an example.
The cubic vertex of noncommutative Chern-Simons theory
consists of a product of $A_0$, $A_1$, and $A_2$.
Therefore, one of the three contractions in Figure 2
must be $\vev{A_2 A_2}$.

%%%%% Figure 2 %%%%%
\begin{figure}
\centerline{\epsfxsize=2in\epsfbox{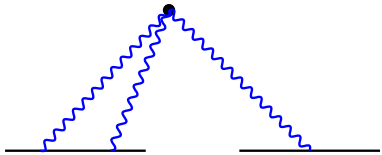}}
\caption{Vanishing interaction graph.}
\end{figure}
%%%%%%%%%%%%%%%%%%%%

Now consider diagrams with internal loops.
To lowest nontrivial order, $O(g^4)$,
this corresponds to the diagrams in Figure 3.\footnote
{The second diagram involves the ghost loop arising from
the usual gauge fixing of the theory, which we have not presented,
and which we do not require in the sequel.}
%%%%% Figure 3 %%%%%
\begin{figure}
\centerline{\epsfxsize=2in\epsfbox{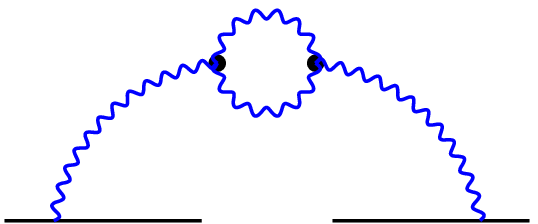} \hspace{1cm}
\epsfxsize=2in\epsfbox{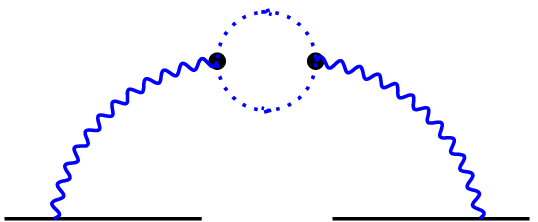} }
\caption{One-loop propagator correction insertions.
The dotted line denotes the ghost propagator.}
\end{figure}
%%%%%%%%%%%%%%%%%%%%
These involve the one-loop corrections to the gauge field propagator.
The calculation of these corrections is similar to those in
commutative {\it non-Abelian} Chern-Simons theory \cite{Chen:2000ak}.
Both diagrams generate the same noncommutative phase structure,
and both can be broken respectively into planar and nonplanar parts
in the standard way \cite{Minwalla:1999px, VanRaamsdonk:2000rr}.
The nonplanar pieces are regulated by the noncommutative phases
\cite{Filk:dm, Minwalla:1999px, VanRaamsdonk:2000rr},\footnote
{See also the paragraph containing (\ref{convergent-integral})
in the next subsection.}
after which the contributions from the ghost loop
and gauge loop rigorously cancel.
On the other hand, the planar pieces of these diagrams,
which are identical to their commutative counterparts
(up to the same overall factor), require careful regularization,
and the study of which ultimately yields the famous one-loop shift
to the Chern-Simons coupling
\cite{Pisarski:1985yj, Alvarez-Gaume:1989wk, Martin:xv}.
However, the one-loop corrections
to the Chern-Simons propagator itself change
neither its tensor structure, nor its momentum dependence.
Thus the arguments of the previous paragraph apply:
the one-loop corrections to
$\vev{A_2(p) A_2(q)}$ still vanish,
and so we conclude that $O(g^4)$ contribution
to $\vev{W(k)W(k')}$ also vanishes:
\begin{equation}
  \vev{W(k)W(k')}
  = (2\pi)^6 \delta^{(3)}(k) \delta^{(3)}(k') + O(g^6).
\end{equation}
This is consistent with the equivalence between
noncommutative Chern-Simons theory
and commutative Chern-Simons theory.

%%%%% Subsection 3.3 %%%%%
\subsection{$\vev{W(k)O_{\mu \nu}(k')}$}

The lowest-order term in $\vev{W(k)O_{\mu \nu}(k')}$
reproduces the result from the commutative side
(\ref{commutative-two-point}) by construction
and corresponds to the diagram in Figure 4.
%%%%% Figure 4 %%%%%
\begin{figure}
\centerline{\epsfxsize=2in\epsfbox{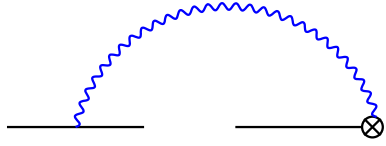}}
\caption{Lowest-order contribution to $\vev{W(k)O_{\mu\nu}(k')}$.
The cross on the right open Wilson line denotes
the field strength insertion.}
\end{figure}
%%%%%%%%%%%%%%%%%%%%
Let us verify this explicitly.
\begin{eqnarray}
  \vev{W(k)O_{\mu \nu}(k')}
  &=& ig (k \theta)^\rho (-i k'_\mu) \vev{A_\rho (k) A_\nu (k')}
    - ( \mu \leftrightarrow \nu ) + O(g^3)
\nonumber \\
  &=& -g (2\pi)^3 \delta^{(3)} (k+k') (k \theta)^\rho k^\sigma
    \frac{k_\mu \epsilon_{\rho \sigma \nu}
           - k_\nu \epsilon_{\rho \sigma \mu}}{k^2} + O(g^3).
\end{eqnarray}
Since the term proportional to $\delta^{(3)}(k)$
in (\ref{SW-map}) is not relevant to the current calculation,
the result (\ref{commutative-two-point}) should be
reproduced by $\vev{W(k)O_{0i}(k')}$
divided by $-g \theta^{12}$. In fact,
\begin{equation}
  -\frac{1}{g \theta^{12}} \vev{W(k)O_{0i}(k')}
  = (2\pi)^3 k_i \delta^{(3)}(k+k') + O(g^2)
  \quad {\rm for} \quad i=1,2.
\end{equation}
Therefore, the question is whether higher-order terms
modify this result or not.
We will calculate $\vev{W(k)O_{\mu \nu}(k')}$ at $O(g^3)$.

Feynman diagrams at $O(g^3)$ fall into two categories.
The first one contains diagrams
which do not have an internal loop.
There are five diagrams in this category, shown in Figures 5 and 6.
The second one contains diagrams
involving the one-loop correction to the propagator,
which are displayed in Figure 7.

Calculations of the five diagrams
in the first category are given in Appendix B.
Here we only present the final results.
%%%%% Figure 5 %%%%%
\begin{figure}
\centerline{\epsfxsize=2in\epsfbox{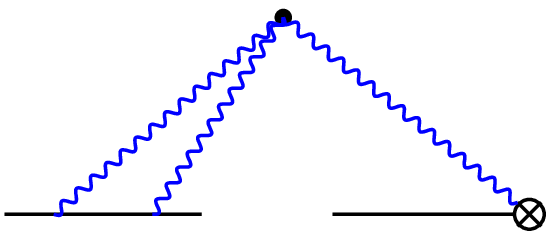} \hspace{1cm}
\epsfxsize=2in\epsfbox{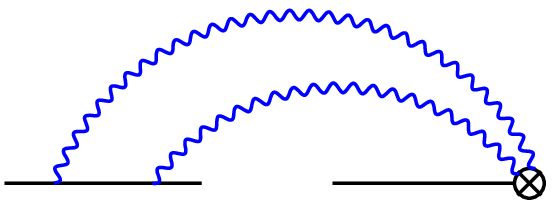}  }
\caption{Diagrams 1 and 2. Canceling $O(g^3)$ corrections to
$\vev{W(k)O_{\mu \nu}(k')}$ involving two gauge field sources
on the pure open Wilson line.
}
\end{figure}
%%%%%%%%%%%%%%%%%%%%

\noindent
{\bf Diagram 1}
\begin{eqnarray}
  && \vev{W(k) O_{\mu \nu}(k')}_{\rm Diagram~1}
\nonumber \\
  && = -(ig)^3 \delta^{(3)} (k+k')
    \int_\epsilon^{1-\epsilon} d \sigma
    \int d^3 p~
    \frac{e^{-i k \times p \sigma}}{p^2 (k-p)^2}
\nonumber \\ && \qquad \qquad \times
    \left\{ (k \times p) (k_\mu l_\nu -  k_\nu l_\mu)
             - l^2 (k_\mu p_\nu - k_\nu p_\mu) \right\}.
\label{diagram-1}
\end{eqnarray}

\noindent
{\bf Diagram 2}
\begin{eqnarray}
  && \vev{W(k) O_{\mu \nu}(k')}_{\rm Diagram~2}
\nonumber \\
  && = (ig)^3 \delta^{(3)} (k+k')
    \int_0^{1-2 \epsilon} d \sigma
    \int d^3 p~
    \frac{e^{-i k \times p \sigma}}
         {p^2 (k-p)^2}
\nonumber \\ && \qquad \qquad \times
    \left\{
    (k \times p) (k_\mu l_\nu - k_\nu l_\mu)
    - l^2 (k_\mu p_\nu - k_\nu p_\mu) \right\}.
\end{eqnarray}

%%%%% Figure 6 %%%%%
\begin{figure}
\centerline{\epsfxsize=1.9in\epsfbox{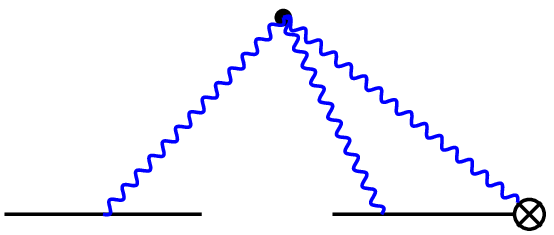} \hspace{0.1cm}
\epsfxsize=1.9in\epsfbox{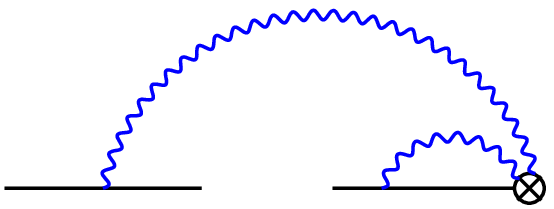} \hspace{0.1cm}
\epsfxsize=1.9in\epsfbox{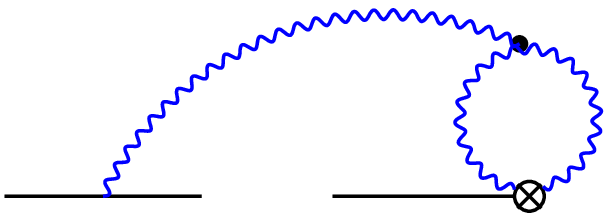} }
\caption{Diagrams 3, 4 and 5.  Canceling $O(g^3)$ corrections to
$\vev{W(k)O_{\mu \nu}(k')}$ involving one gauge field source
on the pure open Wilson line.}
\end{figure}
%%%%%%%%%%%%%%%%%%%%
\noindent
{\bf Diagram 3}
\begin{eqnarray}
  && \vev{W(k) O_{\mu \nu}(k')}_{\rm Diagram~3}
\nonumber \\
  && = 2 (ig)^3 \delta^{(3)} (k+k')
    \int_\epsilon^{1-\epsilon} d \sigma
    \int d^3 p~ e^{i k \times p \sigma}~
    \frac{l^2 (k_\mu p_\nu - k_\nu p_\mu)}{k^2 p^2}
\nonumber \\ && \qquad
  -i (ig)^3 \delta^{(3)} (k+k')
    \int d^3 p~
    \frac{ e^{i k \times p (1-\epsilon)}
           - e^{i k \times p \epsilon} }
         {k^2 p^2 (k+p)^2}
\nonumber \\ && \qquad \quad \times
  \biggl[ -(k^2 + k \cdot p) ( k_\mu l_\nu - k_\nu l_\mu)
  - (k^2 + 2 k \cdot p) ( p_\mu l_\nu - p_\nu l_\mu)
\nonumber \\ && \qquad \qquad \quad
  + 2 (k \times p) (p_\mu k_\nu  - p_\nu k_\mu) \biggr].
\end{eqnarray}

\noindent
{\bf Diagram 4}
\begin{eqnarray}
  && \vev{W(k) O_{\mu \nu}(k')}_{\rm Diagram~4}
\nonumber \\
  && = -2 (ig)^3 \delta^{(3)} (k+k')
  \int_{\epsilon}^{1-2 \epsilon} d \sigma
  \int d^3 p~
  e^{i k \times p \sigma}~
  \frac{l^2 (k_\mu p_\nu - k_\nu p_\mu)}{k^2 p^2}
\nonumber \\ && \quad
  -2i (ig)^3 \delta^{(3)} (k+k')
  \int d^3 p~
  \left\{ e^{i k \times p (1 - 2 \epsilon)}
          - e^{i k \times p \epsilon} \right\}
  \frac{k_\mu l_\nu - k_\nu l_\mu}{k^2 p^2}.
\end{eqnarray}

\noindent
{\bf Diagram 5}
\begin{eqnarray}
  && \vev{W(k) O_{\mu \nu}(k')}_{\rm Diagram~5}
\nonumber \\
  && = i (ig)^3 \delta^{(3)} (k+k')
    \int d^3 p~
    \frac{ e^{i k \times p (1-\epsilon)}
           - e^{i k \times p \epsilon} }
         {k^2 p^2 (k+p)^2}
\nonumber \\ && \qquad \times
   \biggl[
   -2 (k \times p) (k_\mu p_\nu - k_\nu p_\mu)
   -(2 p^2 + k \cdot p) (l_\mu k_\nu - l_\nu k_\mu)
\nonumber \\ && \qquad \qquad
   +(k^2 + 2 k \cdot p) (l_\mu p_\nu - l_\nu p_\mu)
   \biggr].
\label{diagram-5}
\end{eqnarray}

Let us first consider whether or not
each of the five contributions is finite.
All of the integrals over the momentum $p$ take
the following form:
\begin{equation}
  \int d^3 p~ f(p_0, p_1, p_2)~ e^{i k \times p \sigma},
\label{convergent-integral}
\end{equation}
where $f(p_0, p_1, p_2)$ is
a meromorphic function of $p_0$, $p_1$, and $p_2$.
If the $\theta$-dependent phase factor is absent,
the integral can be divergent.
However, as is well-known
\cite{Minwalla:1999px, VanRaamsdonk:2000rr},
the phase factor makes the integral convergent.
Let us illustrate this point in the simple example where
$f(p_0, p_1, p_2) = 1/[(2 \pi)^3 (p_0^2 + p_1^2 + p_2^2)]$.
We can choose a coordinate system such that $k_2=0$.
The integral becomes
\begin{equation}
  \int \frac{d^3 p}{(2 \pi)^3}~
  \frac{e^{i k_1 \theta^{12} p_2 \sigma}}{p_0^2 + p_1^2 + p_2^2}.
\end{equation}
The integral over $p_2$ can be carried out
by evaluating the residue of the pole at
either $p_2 = i \sqrt{p_0^2+p_1^2}$
or $p_2 = -i \sqrt{p_0^2+p_1^2}$
depending on the sign of $k_1 \theta^{12} \sigma$.
The remaining integrals over $p_0$ and $p_1$ are
easily performed to give
\begin{equation}
  \int_{-\infty}^{\infty} \frac{d p_0}{2 \pi}
  \int_{-\infty}^{\infty} \frac{d p_1}{2 \pi}~
  \frac{e^{- | k_1 \theta^{12} \sigma | \sqrt{p_0^2+p_1^2}}}
       {2 \sqrt{p_0^2+p_1^2}}
  = \frac{1}{4 \pi | k_1 \theta^{12} \sigma |}.
\end{equation}
This is nothing but the calculation of Green's function
in three dimensions
if we replace $(k \theta)^\mu \sigma$ by $x^\mu$.
Note that the integral over $p_2$ provided
a damping factor which exponentially suppresses
the integrand for large $p$ and thus makes
the integrals over $p_0$ and $p_1$ converge.
This mechanism works in general as long as
the phase factor is nonvanishing.

Therefore, the integrals over $p$ in the five diagrams
converge as long as $\sigma$ is nonzero.
Only the integral in Diagram 2 is potentially dangerous,
but we can show that it is convergent as well.
What we need to show is the following:
\begin{equation}
    \lim_{\epsilon \to 0}
    \int_0^{\epsilon} d \sigma
    \int d^3 p~
    \frac{e^{-i k \times p \sigma}}
         {p^2 (k-p)^2}
    \left\{
    (k \times p) (k_\mu l_\nu - k_\nu l_\mu)
    - l^2 (k_\mu p_\nu - k_\nu p_\mu) \right\} = 0.
\label{finiteness}
\end{equation}
For the first term, only the surface terms of
the integral over $\sigma$ contribute:
\begin{equation}
    i (k_\mu l_\nu - k_\nu l_\mu)
    \int d^3 p~
    \frac{e^{-i k \times p \epsilon}-1}
         {p^2 (k-p)^2}.
\end{equation}
Now the term coming from $\sigma=0$ is also finite
by power counting
so that we can safely take the limit $\epsilon \to 0$.
The calculation of the second term reduces to
\begin{equation}
    \int d^3 p~
    \frac{p_\mu~ e^{-i k_1 \theta^{12} p_2 \sigma}}
         {(\frac{k}{2} - p)^2 (\frac{k}{2} + p)^2},
\end{equation}
where we have changed variables as $p'=k/2-p$
and chosen a coordinate system such that $k_2=0$ as usual.
When $p_\mu = p_2$, the calculation reduces to
the case of the first term in (\ref{finiteness}).
When $p_\mu = p_0$ or $p_\mu = p_1$,
the integral vanishes for a nonzero $\sigma$
because the integrand after the integral over $p_2$
is odd in $(p_0, p_1) \to (-p_0, -p_1)$.
Thus we have shown (\ref{finiteness})
and confirmed that each of the contributions coming from
the five diagrams is finite.

The contributions from Diagram 1 and Diagram 2 almost cancel.
The sum of the two can be written as follows:
\begin{eqnarray}
    && (ig)^3 \delta^{(3)} (k+k')
    \Biggl[
    \int_0^{\epsilon} d \sigma
    \int d^3 p~
    \frac{e^{-i k \times p \sigma}}
         {p^2 (k-p)^2}
    \left\{
    (k \times p) (k_\mu l_\nu - k_\nu l_\mu)
    - l^2 (k_\mu p_\nu - k_\nu p_\mu) \right\}
\nonumber \\
    && \qquad
    - \int_{1-2 \epsilon}^{1- \epsilon} d \sigma
    \int d^3 p~
    \frac{e^{-i k \times p \sigma}}
         {p^2 (k-p)^2}
    \left\{
    (k \times p) (k_\mu l_\nu - k_\nu l_\mu)
    - l^2 (k_\mu p_\nu - k_\nu p_\mu) \right\} \Biggr].
\end{eqnarray}
As we have seen,
the first term vanishes in the limit $\epsilon \to 0$.
The second term is less dangerous and also vanishes in the limit.
We thus conclude that the contributions from Diagram 1
and Diagram 2 cancel.

Now consider the remaining three diagrams.
The contributions from Diagram 3 and Diagram 4 contain
integrals over $\sigma$.
The two integrals almost cancel and the difference
vanishes in the limit $\epsilon \to 0$ as before.
The remaining terms which do not contain
an integral over $\sigma$ share a similar structure.
The term in Diagram 4 is different
in that it contains $e^{i k \times p (1 - 2 \epsilon)}$.
However, we can replace it by $e^{i k \times p (1 - \epsilon)}$
since the difference vanishes in the limit $\epsilon \to 0$:
\begin{equation}
  \lim_{\epsilon \to 0}
  \int d^3 p~
  \left\{ e^{i k \times p (1 - 2 \epsilon)}
          - e^{i k \times p (1 - \epsilon)} \right\}
  \frac{k_\mu l_\nu - k_\nu l_\mu}{k^2 p^2} = 0.
\end{equation}
Now the sum of the terms from the three diagrams
can be written as follows:
\begin{eqnarray}
    && -i (ig)^3 \delta^{(3)} (k+k')
    \int d^3 p~
    \frac{ e^{i k \times p (1-\epsilon)}
           - e^{i k \times p \epsilon} }
         {k^2 p^2 (k+p)^2}
   (k^2 + 2 k \cdot p) ( k_\mu l_\nu - k_\nu l_\mu)
\nonumber \\
  && = -i (ig)^3 \delta^{(3)} (k+k')
    \int d^3 p~
    \frac{ e^{i k \times p (1-\epsilon)}
           - e^{i k \times p \epsilon} }
         {k^2}
    \left[ \frac{1}{p^2} - \frac{1}{(k+p)^2} \right]
    ( k_\mu l_\nu - k_\nu l_\mu)
  = 0,
\nonumber \\
\end{eqnarray}
where we have changed variables as $p'=p+k$
for the term involving $1/(k+p)^2$ in the last line.
To summarize, we have seen that in the limit $\epsilon \to 0$
the contributions from Diagram 1 and Diagram 2 cancel,
and those from Diagram 3, 4, and 5 cancel among themselves.

The second category of diagrams at $O(g^3)$, displayed
in Figure 7, contain the one-loop correction to the propagator. 
%%%%% Figure 7 %%%%%
\begin{figure}
\centerline{\epsfxsize=2in\epsfbox{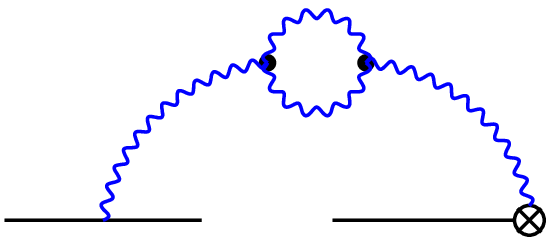} \hspace{1cm}
\epsfxsize=2in\epsfbox{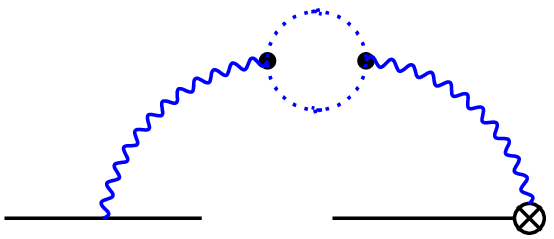} }
\caption{$O(g^3)$ contributions to $\vev{W(k)O_{\mu\nu}(k')}$
containing propagator corrections.}
\end{figure}
%%%%%%%%%%%%%%%%%%%%
As we have discussed in the previous subsection,
the nonplanar contributions from these diagrams
are finite and cancel
between the gauge-field and ghost diagrams,
and the planar contributions only renormalize
the overall coefficient of the tree-level result:
\begin{equation}
  -\frac{1}{g \theta^{12}} \vev{W(k)O_{0i}(k')}
  = \left[ 1 + O(g^2) \right]
  (2\pi)^3 k_i \delta^{(3)}(k+k') + O(g^4)
  \quad {\rm for} \quad i=1,2.
\end{equation}
Does this violate the equivalence between
noncommutative and commutative Chern-Simons theories?

The disagreement is coming from
different wave-function renormalizations
between the commutative and noncommutative theories.
The commutative $U(1)$ Chern-Simons theory is free
and its propagator $\vev{a_\mu (p) a_\nu (q)}$
does not receive any wave-function renormalization.
On the other hand,
the tree-level propagator $\vev{A_\mu (p) A_\nu (q)}$
in the noncommutative theory can be renormalized
by quantum effects depending on a regularization scheme,
and this is precisely the origin of the one-loop correction
to $\vev{W(k)O_{\mu \nu}(k')}$.
However, since the correction changes
neither the tensor structure nor the momentum dependence
and only modifies the overall coefficient,
its effect can be absorbed in renormalizations
of the composite operators $W(k)$ and $O_{\mu \nu}(k)$.
Therefore, the equivalence between the commutative
and noncommutative theories still holds if we modify
the Seiberg-Witten map at the classical level (\ref{SW-map}) to
\begin{eqnarray}
  f_{12}(k)
  &=& -\frac{Z}{g \theta^{12}}
    \left[ W(k) - (2\pi)^3 \delta^{(3)}(k) \right],
\nonumber \\
  f_{0i}(k) &=& Z O_{0i}(k) \quad {\rm for} \quad i=1,2,
\label{quantum-SW-map}
\end{eqnarray}
such that the renormalization factor $Z$ compensates for
the correction to the two-point function
$\vev{W(k)O_{\mu \nu}(k')}$.
Note that the two renormalization factors in (\ref{quantum-SW-map})
must be the same in order for the Bianchi identity to hold.

It is not too surprising that we found a scheme-dependent
quantum correction to the Seiberg-Witten map
because the Seiberg-Witten map is a special kind of
field redefinition between the commutative variable $a_\mu$
and the noncommutative variable $A_\mu$,
and these fields in general
suffer from scheme-dependent wave-function renormalizations.
Our expression for the Seiberg-Witten map was designed
to satisfy the initial condition (\ref{initial-condition})
classically, but a quantum correction is necessary
if the wave-function renormalization is singular
in the limit $\theta \to 0$
as it is in the case of noncommutative Chern-Simons theory.

%%%%% Subsection 3.4 %%%%%
\subsection{$\vev{W(k_1)W(k_2)W(k_3)}$}
Let us calculate the three-point function
of pure open Wilson lines.
This should vanish except for the trivial lowest-order term
in order for the correspondence between the commutative
and noncommutative sides to hold.

The first nontrivial contribution starts at $O(g^4)$.
Let us expand $\vev{W(k_1)W(k_2)W(k_3)}$
up to $O(g^4)$:
\begin{eqnarray}
  && \vev{W(k_1)W(k_2)W(k_3)}
\nonumber \\
  && = (2 \pi)^9 \delta^{(3)}(k_1) \delta^{(3)}(k_2)
                                   \delta^{(3)}(k_3)
\nonumber \\ && \qquad
  {}+ (ig)^3 \vev{l_1 \cdot A(k_1)~ l_2 \cdot A(k_2)~
                                  l_3 \cdot A(k_3)}
\nonumber \\ && \qquad
  {}+ \frac{(ig)^4}{2} \int_\epsilon^{1-\epsilon} d \sigma
    \int \frac{d^3 p}{(2\pi)^3}~
\nonumber \\ && \qquad \qquad \times
    \Bigl[
    e^{-i k_3 \times p \sigma + \frac{i}{2} k_3 \times p}
    \vev{l_1 \cdot A(k_1)~ l_2 \cdot A(k_2)~
         l_3 \cdot A(p)~ l_3 \cdot A(k_3-p)}
\nonumber \\ && \qquad \qquad \qquad
    + ( (k_1, k_2, k_3) \to (k_2, k_3, k_1) )
    + ( (k_1, k_2, k_3) \to (k_3, k_1, k_2) ) \Bigr]
\nonumber \\ && \qquad
    {}+ O(g^5),
\end{eqnarray}
where
$l_i^\mu \equiv (k_i \theta)^\mu = (k_i)_\nu \theta^{\nu \mu}$
for $i=1,2,3$.
The term at $O(g^3)$ contracted with the vertex (\ref{vertex})
and the terms at $O(g^4)$
contracted with two propagators of (\ref{propagator})
contribute at $O(g^4)$.
These correspond to the diagrams in Figure 8.
%%%%% Figure 8 %%%%%
\begin{figure}
\centerline{
\epsfxsize=1.8in\epsfbox{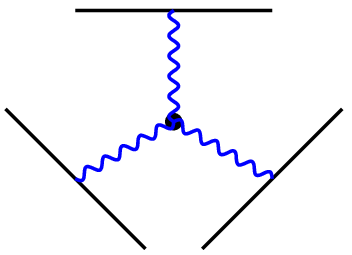} \hspace{1cm}
\epsfxsize=1.8in\epsfbox{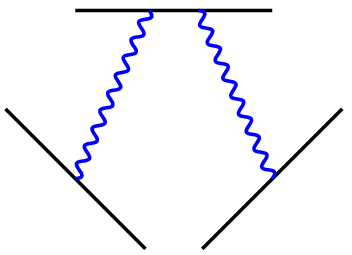} }
\caption{$O(g^4)$ contributions to $\vev{W(k_1)W(k_2)W(k_3)}$.}
\end{figure}
%%%%%%%%%%%%%%%%%%%%

Let us begin with the latter.
There are two nonvanishing contractions
for each of the three terms at $O(g^4)$.
The two contractions are combined
to give the following expression:
\begin{eqnarray}
    && \frac{(ig)^4}{2} \int_\epsilon^{1-\epsilon} d \sigma
    \int \frac{d^3 p}{(2\pi)^3}~
    e^{-i k_3 \times p \sigma + \frac{i}{2} k_3 \times p}
\nonumber \\ && \qquad \qquad \times
    \vev{l_1 \cdot A(k_1)~ l_2 \cdot A(k_2)~
         l_3 \cdot A(p)~ l_3 \cdot A(k_3-p)}
\nonumber \\ &&
    = (ig)^4 (2 \pi)^3 \delta^{(3)} (k_1+k_2+k_3)
    \int_\epsilon^{1-\epsilon} d \sigma~
    \cos \left[ k_1 \times k_2 \left(
    \sigma - \frac{1}{2} \right) \right]
\nonumber \\ && \qquad \qquad \times
    l_1^{\mu_1} l_2^{\mu_2} l_3^{\nu_1} l_3^{\nu_2}
    \epsilon_{\mu_1 \rho_1 \nu_1} \epsilon_{\mu_2 \rho_2 \nu_2}
    \frac{k_1^{\rho_1}}{k_1^2} \frac{k_2^{\rho_2}}{k_2^2}.
\end{eqnarray}
Since the vectors $l_i^\mu$ have vanishing 0-components,
the indices $\mu_1$, $\nu_1$, $\mu_2$, and $\nu_2$ cannot be zero.
Therefore, the indices $\rho_1$ and $\rho_2$ must be zero
in order for the expression to be nonvanishing.
If we decompose the vectors $k_i$ as
\begin{equation}
  k_i = (\omega_i, \vec{k}_i)
\end{equation}
for $i=1,2,3$
where $\omega_i = (k_i)^0$ and ($\vec{k}_i)^\mu = (k_i)^\mu$
for $\mu = 1, 2$, we have
\begin{eqnarray}
  && l_1^{\mu_1} l_2^{\mu_2} l_3^{\nu_1} l_3^{\nu_2}
    \epsilon_{\mu_1 \rho_1 \nu_1} \epsilon_{\mu_2 \rho_2 \nu_2}
    \frac{k_1^{\rho_1}}{k_1^2} \frac{k_2^{\rho_2}}{k_2^2}
\nonumber \\
  && = \frac{(\theta^{12})^2 \omega_1 \omega_2
          (k_1 \times k_3)(k_2 \times k_3)}{k_1^2 k_2^2}
  = -\frac{(\theta^{12})^2 \omega_1 \omega_2 (k_1 \times k_2)^2}
          {k_1^2 k_2^2},
\end{eqnarray}
where we have used
\begin{equation}
  l_i^\mu \epsilon_{\mu 0 \nu} l_j^\nu
  = -\theta^{12} (k_i \times k_j),
\end{equation}
and $k_1 \times k_2 = k_2 \times k_3 = k_3 \times k_1$
by momentum conservation.
The integral over $\sigma$ is straightforward
and we do not have any singularity when we take $\epsilon \to 0$.
The result is thus given by
\begin{eqnarray}
    -2 (ig)^4 (2 \pi)^3 \delta^{(3)} (k_1+k_2+k_3)
    \sin \left( \frac{k_1 \times k_2}{2} \right)
    \frac{(\theta^{12})^2 k_3^2 \omega_1 \omega_2 (k_1 \times k_2)}
         {k_1^2 k_2^2 k_3^2}.
\end{eqnarray}

We also need to add the two terms coming from the permutations
$(k_1, k_2, k_3) \to (k_2, k_3, k_1)$
and $(k_1, k_2, k_3) \to (k_3, k_1, k_2)$.
Since $k_1^2 k_2^2 k_3^2$ is invariant under the permutations
and $k_1 \times k_2 = k_2 \times k_3 = k_3 \times k_1$,
the only nontrivial part is $k_3^2 \omega_1 \omega_2$.
We can eliminate $\omega_3$ and $\vec{k}_3$
using momentum conservation to find
\begin{eqnarray}
  && k_3^2 \omega_1 \omega_2
  + k_1^2 \omega_2 \omega_3 + k_2^2 \omega_3 \omega_1
\nonumber \\
  && = 2 \omega_1 \omega_2 \vec{k}_1 \cdot \vec{k}_2
    - \omega_2^2 \vec{k}_1^2 - \omega_1^2 \vec{k}_2^2
  = \frac{2 \omega_1 \omega_2 l_1 \cdot l_2
          - \omega_2^2 l_1^2 - \omega_1^2 l_2^2 }
         {(\theta^{12})^2},
\end{eqnarray}
where we have used
\begin{equation}
  \vec{k}_i \cdot \vec{k}_j
  = \frac{1}{(\theta^{12})^2} l_i \cdot l_j.
\label{l-and-vector}
\end{equation}
Therefore, the contribution to $\vev{W(k_1)W(k_2)W(k_3)}$
at $O(g^4)$ coming from the sum of the diagrams
without cubic vertices is given by
\begin{eqnarray}
    2 (ig)^4 (2 \pi)^3 \delta^{(3)} (k_1+k_2+k_3)
    \sin \left( \frac{k_1 \times k_2}{2} \right)
    \frac{( -2 \omega_1 \omega_2 l_1 \cdot l_2
            + \omega_2^2 l_1^2 + \omega_1^2 l_2^2 )
          (k_1 \times k_2)}
         {k_1^2 k_2^2 k_3^2}.
\nonumber \\
\label{3-point-1}
\end{eqnarray}

Let us next calculate the contribution from the diagram
with a cubic vertex.
It can be evaluated using (\ref{vertex})
and (\ref{4-epsilons}) as follows:
\begin{eqnarray}
  && (ig)^3 \vev{l_1 \cdot A(k_1)~ l_2 \cdot A(k_2)~
                                   l_3 \cdot A(k_3)}
\nonumber \\ &&
  = -2 (ig)^4 (2 \pi)^3 \delta^{(3)} (k_1+k_2+k_3)
    \sin \left( \frac{k_1 \times k_2}{2} \right)
\nonumber \\ && \qquad \times
    \frac{-2 (k_1 \times k_2)^3
          + (k_1 \times k_2) \{k_1^2 l_2^2 + k_2^2 l_1^2
                               -2 (k_1 \cdot k_2) (l_1 \cdot l_2)\} }
         {k_1^2 k_2^2 k_3^2}.
\label{3-point-2}
\end{eqnarray}
Apparently, this does not seem to cancel the contribution
(\ref{3-point-1}). For example, the expression (\ref{3-point-1})
vanishes when $\omega_1=\omega_2=0$, but it is not obvious
that this also holds in (\ref{3-point-2}).
Let us take a closer look at the factor
$k_1^2 l_2^2 + k_2^2 l_1^2 -2 (k_1 \cdot k_2) (l_1 \cdot l_2)$.
It can be decomposed in the following way:
\begin{eqnarray}
  && k_1^2 l_2^2 + k_2^2 l_1^2 -2 (k_1 \cdot k_2) (l_1 \cdot l_2)
  = (\omega_1^2 + \vec{k}_1^2) l_2^2
     + (\omega_2^2 + \vec{k}_2^2) l_1^2
     -2 (\omega_1 \omega_2 + \vec{k}_1 \cdot \vec{k}_2)
         l_1 \cdot l_2
\nonumber \\ &&
  = \omega_1^2 l_2^2 + \omega_2^2 l_1^2
    -2 \omega_1 \omega_2 l_1 \cdot l_2
    + 2 (\theta^{12})^2 \{ \vec{k}_1^2 \vec{k}_2^2
                           - (\vec{k}_1 \cdot \vec{k}_2)^2 \}
\nonumber \\ &&
  = \omega_1^2 l_2^2 + \omega_2^2 l_1^2
    -2 \omega_1 \omega_2 l_1 \cdot l_2
    + 2 (k_1 \times k_2)^2,
\end{eqnarray}
where we have used (\ref{l-and-vector}) and
\begin{equation}
  (k_i \times k_j)^2 = (\vec{k}_i \times \vec{k}_j)^2
  = (\theta^{12})^2 \{ \vec{k}_i^2 \vec{k}_j^2
                       - (\vec{k}_i \cdot \vec{k}_j)^2 \}.
\end{equation}
Therefore, we have
\begin{eqnarray}
  && (ig)^3 \vev{l_1 \cdot A(k_1)~ l_2 \cdot A(k_2)~
                                   l_3 \cdot A(k_3)}
\nonumber \\ &&
  = -2 (ig)^4 (2 \pi)^3 \delta^{(3)} (k_1+k_2+k_3)
    \sin \left( \frac{k_1 \times k_2}{2} \right)
\nonumber \\ && \qquad \times
    \frac{ (k_1 \times k_2) (\omega_1^2 l_2^2 + \omega_2^2 l_1^2
             -2 \omega_1 \omega_2 l_1 \cdot l_2)}
         {k_1^2 k_2^2 k_3^2}.
\end{eqnarray}
This precisely cancels (\ref{3-point-1}) so that
the sum of all the contributions to
the three-point function
$\vev{W(k_1)W(k_2)W(k_3)}$ at $O(g^4)$ vanishes. Thus
\begin{equation}
  \vev{W(k_1)W(k_2)W(k_3)}
  = (2 \pi)^9 \delta^{(3)}(k_1) \delta^{(3)}(k_2)
              \delta^{(3)}(k_3) + O(g^5).
\end{equation}
This is again consistent with the equivalence between
noncommutative and commutative Chern-Simons theories.

%%%%% Section 4 %%%%%
\section{Conclusions and discussion}
\setcounter{equation}{0}

We have calculated the two-point functions
$\vev{W(k) W(k')}$ and $\vev{W(k) O_{\mu \nu}(k')}$,
and the three-point function $\vev{W(k_1) W(k_2) W(k_3)}$
in noncommutative Chern-Simons theory,
and compared them with their commutative counterparts.
We found the equivalence between
commutative and noncommutative Chern-Simons theories
with respect to these observables
persists at the first nontrivial order in perturbation theory.

The agreement in the two-point functions
may seem more or less trivial
since the topological nature
of the theory strongly constrains the possible form
of the correlation functions.
In practice, however, we need to choose a gauge, which
inevitably introduces metric dependence, and introduce
a regulator to make the computation well-defined.
We have acquired insight into interesting quantum aspects
of the Seiberg-Witten map from the computation.
First, the relation between the regularizations
of the integral over $\sigma$ and the commutator,
which is closely connected with the Bianchi identity
as we discussed in Subsection 2.2,
did play an important role in the cancellations
we found in the calculation of $\vev{W(k) O_{\mu \nu}(k')}$
in Subsection 3.3.
Another interesting aspect we have encountered in the calculation
is the quantum correction to the Seiberg-Witten map
(\ref{quantum-SW-map}).
These seem to provide us with some insight into
how we should define
the composite operators $W(k)$ and $O_{\mu \nu}(k)$
at the quantum level.

The agreement in the three-point function is more nontrivial.
Although there is no dependence on the metric,
$W(k)$ depends on $\theta^{\mu \nu}$ as well as its momentum
so that the three-point function
could be a nontrivial function of $k_1 \times k_2$.
The nontrivial dependence on $k_1 \times k_2$
is not excluded by the topological nature of the theory,
while the equivalence to the commutative theory
requires it to vanish.
We did find that
it vanishes at the first nontrivial order in $g$.\footnote
{It was argued in \cite{Das:2001kf, Martin:2001ci} that
noncommutative Chern-Simons theory is a free theory from
an analysis in the axial gauge.
We would like to comment on the relation between
this work and ours.
First of all, correlation functions of composite operators,
such as $W(k)$ or $O_{\mu \nu}(k)$ in our case, are in general
nontrivial even in a free theory.
For example, a vacuum expectation value of a Wilson loop
is nontrivial in the free Abelian $F^2$ gauge theory.
Therefore, our results are not immediate consequences
of the observations in \cite{Das:2001kf, Martin:2001ci}.
Technically, however, our calculations could have been
much simplified in the axial gauge.
Although it was argued in \cite{Das:2001kf, Martin:2001ci} that
the axial gauge can be safely taken
in noncommutative Chern-Simons theory,
calculations involving open Wilson lines can be subtle
because it is essential that
we are able to perform integration by parts in proving
various properties of open Wilson lines, such as the Bianchi
identity, while the propagator does not decay at infinity
in the axial gauge.
For example, $W(k)$ with $k_2=0$ becomes trivial
in the gauge $A_2=0$, but it is inconsistent with
our result in the covariant gauge where
$\vev{W(k) O_{0i}(k')}$ is nonvanishing.}

It would be an interesting question
as to whether or not the equivalence between
the commutative and noncommutative Chern-Simons theories
persists to higher orders in $g$ \cite{Kaminsky},
or even nonperturbatively.\footnote
{Rigorously speaking, pure noncommutative Chern-Simons theory
without any additional degrees of freedom has not been realized
in string theory.
The realizations given in \cite{Bergman:2001qg, Hellerman:2001yv}
contain additional degrees of freedom
which are important for quasi-particle or quasi-hole excitations
of quantum Hall fluid, or to realize quantum Hall fluid
with a boundary.
Therefore, the equivalence between the commutative
and noncommutative theories is not guaranteed
by its embedding in string theory.}
It has been noted
in \cite{Sheikh-Jabbari:2001au, Nair:2001rt, Bak:2001ze}
that the level of noncommutative Chern-Simons theory
is quantized even for the $U(1)$ gauge group,
while that for the commutative theory is not
because of the difference in the gauge group topologies
of the two cases \cite{Harvey:2001pd}.
This raises a question on the equivalence of the two theories
at the nonperturbative level.\footnote{We would like to thank
D.~Bak, M.~M.~Sheikh-Jabbari, and A.~P.~Polychronakos
for drawing our attention to this point.}
Clearly our perturbative
computation does not address this issue,
leaving this as an interesting future problem.
If the equivalence holds nonperturbatively
under the Seiberg-Witten map (\ref{SW-map})
up to possible quantum corrections to the map itself,
correlation functions of $W(k)$ and $O_{\mu \nu}(k)$
are rather trivial in the sense that
they are exactly given by their commutative counterparts.
It would be an interesting future problem
to construct more nontrivial observables, if any,
in noncommutative Chern-Simons theory in this case.

Finally, let us discuss the generalization of our results
to the $U(N)$ case.
The map (\ref{SW-map}) between gauge-invariant observables
on the commutative and noncommutative sides
can be easily generalized to the $U(N)$ case
by studying the coupling of multiple D-branes
to the Ramond-Ramond potentials with a constant $B$ field.
The map in the case of $U(N)$ is simply given by
taking the trace of (\ref{SW-map}):
\begin{eqnarray}
  {\rm tr}~ f_{12}(k)
  &=& -\frac{1}{g \theta^{12}}
    {\rm tr} \left[ W(k) - (2\pi)^3 \delta^{(3)}(k)
{\bf 1} \right],
\nonumber \\
  {\rm tr}~ f_{0i}(k)
  &=& {\rm tr}~ O_{0i}(k) \quad {\rm for} \quad i=1,2,
\label{U(N)-map}
\end{eqnarray}
where ${\bf 1}$ is the identity matrix.
A Feynman diagram of correlation functions
involving ${\rm tr}~ W(k)$ and ${\rm tr}~ O_{\mu \nu}(k)$
can be evaluated by multiplying the contribution from
the same diagram in the $U(1)$ case
by an appropriate power of $N$.

All five diagrams of $\vev{W(k) O_{\mu \nu}(k')}$
at $O(g^3)$ not containing an internal loop
scale as $N^2$ so that the cancellations we found
remain intact.
The two diagrams displayed in Figure 8
for $\vev{W(k_1)W(k_2)W(k_3)}$
at $O(g^4)$ scale as $N$. Therefore, the three-point function
of ${\rm tr}~ W(k)$ also vanishes at $O(g^4)$ in the $U(N)$ case.
As for the one-loop corrections to the propagator,
the cancellation of the nonplanar pieces persists
and only the planar pieces produce a nonvanishing contribution
just as in the case of $U(1)$.
We should note that one-loop corrections
to the propagator also exist on the commutative side
in the case of $U(N)$
\cite{Pisarski:1985yj, Alvarez-Gaume:1989wk, Martin:xv}.
If we use the same regularization on the commutative
and noncommutative sides, we have the same one-loop corrections
on both sides so that one-loop corrections
to the map (\ref{U(N)-map}) are absent in the $U(N)$ case.
We thus conclude that the equivalence between $U(N)$ noncommutative
and commutative Chern-Simons theories also holds at the first
nontrivial order in $g$
under the map (\ref{U(N)-map}) between gauge-invariant observables.

%%%%% Acknowledgements %%%%%
\section*{Acknowledgements}
Y.O. and H.O. would like to thank
the Aspen Center for Physics, where part of this work was done.
This work was supported in part by the DOE grant DE-FG03-92ER40701.
The work of K.K. was supported in part by
the Natural Sciences and Engineering Research Council of Canada.
The work of Y.O. was supported in part
by a John A. McCone Fellowship in Theoretical Physics from
California Institute of Technology.

%\newpage

%%%%%%%%%% Appendices %%%%%%%%%%
\appendix
\renewcommand{\thesection}{Appendix \Alph{section}.}
\renewcommand{\theequation}{\Alph{section}.\arabic{equation}}

%%%%% Appendix A %%%%%
\section{Conventions and Feynman rules}
\setcounter{equation}{0}
The action of noncommutative Chern-Simons theory
in terms of a canonically normalized gauge field is given by
\begin{equation}
  S_{NCCS} = \frac{1}{2} \int d^3 x~ \epsilon^{\mu \rho \nu}
  \left[
    A_\mu \ast \partial_\rho A_\nu
    - \frac{2ig}{3} A_\mu \ast A_\rho \ast A_\nu
  \right].
\end{equation}
Our Fourier transform convention is
\begin{equation}
  A_\mu (x) = \int \frac{d^3 p}{(2\pi)^3}~
              e^{-ikx} A_\mu (k).
\end{equation}
We use the standard covariant gauge-fixing term proportional to
$(\partial \cdot A)^2$ and then take the Landau gauge.
The propagator is given by
\begin{equation}
  \vev{A_\mu (p) A_\nu (q)}
  = (2 \pi)^3 \delta^{(3)} (p+q) \epsilon_{\mu \rho \nu}
    \frac{p^\rho}{p^2}.
\label{propagator}
\end{equation}
The three-point function of noncommutative gauge fields
contracted with a cubic vertex is
given by
\begin{eqnarray}
  && \vev{A_{\mu_1} (q_1) A_{\mu_2} (q_2) A_{\mu_3} (q_3) }
\nonumber \\
  && = -2ig (2\pi)^3 \delta^{(3)} (q_1+q_2+q_3)
    \epsilon^{\nu_1 \nu_2 \nu_3}
\nonumber \\ && \qquad \times
    \sin \left( \frac{q_1 \times q_2}{2} \right)
    \frac{\epsilon_{\mu_1 \rho_1 \nu_1} q_1^{\rho_1}}{q_1^2}
    \frac{\epsilon_{\mu_2 \rho_2 \nu_2} q_2^{\rho_2}}{q_2^2}
    \frac{\epsilon_{\mu_3 \rho_3 \nu_3} q_3^{\rho_3}}{q_3^2}.
\label{vertex}
\end{eqnarray}
The following identities are useful:
\begin{eqnarray}
  && \epsilon_{\mu_1 \rho_1 \nu_1} \epsilon_{\mu_2 \rho_2 \nu_2}
    = \delta_{\mu_1 \mu_2} \delta_{\rho_1 \rho_2} \delta_{\nu_1 \nu_2}
    + \delta_{\mu_1 \rho_2} \delta_{\rho_1 \nu_2} \delta_{\nu_1 \mu_2}
    + \delta_{\mu_1 \nu_2} \delta_{\rho_1 \mu_2} \delta_{\nu_1 \rho_2}
\nonumber \\ && \qquad \qquad \qquad \quad
    - \delta_{\mu_1 \mu_2} \delta_{\rho_1 \nu_2} \delta_{\nu_1 \rho_2}
    - \delta_{\mu_1 \nu_2} \delta_{\rho_1 \rho_2} \delta_{\nu_1 \mu_2}
    - \delta_{\mu_1 \rho_2} \delta_{\rho_1 \mu_2} \delta_{\nu_1 \nu_2}.
\label{2-epsilons}
\\
  && \epsilon_{\mu_1 \rho_1 \nu_1} \epsilon_{\mu_2 \rho_2 \nu_2}
     \epsilon_{\mu_3 \rho_3 \nu_3} \epsilon^{\nu_1 \nu_2 \nu_3}
\nonumber \\ && \quad
    = \delta_{\mu_1 \rho_2} \delta_{\rho_1 \rho_3} \delta_{\mu_3 \mu_2}
    + \delta_{\mu_1 \rho_3} \delta_{\rho_1 \mu_2} \delta_{\mu_3 \rho_2}
    + \delta_{\mu_1 \mu_3} \delta_{\rho_1 \rho_2} \delta_{\rho_3 \mu_2}
    + \delta_{\mu_1 \mu_2} \delta_{\rho_1 \mu_3} \delta_{\rho_3 \rho_2}
\nonumber \\ && \qquad
    - \delta_{\mu_1 \mu_2} \delta_{\rho_1 \rho_3} \delta_{\mu_3 \rho_2}
    - \delta_{\mu_1 \rho_3} \delta_{\rho_1 \rho_2} \delta_{\mu_3 \mu_2}
    - \delta_{\mu_1 \rho_2} \delta_{\rho_1 \mu_3} \delta_{\rho_3 \mu_2}
    - \delta_{\mu_1 \mu_3} \delta_{\rho_1 \mu_2} \delta_{\rho_3 \rho_2}.
\label{4-epsilons}
\end{eqnarray}

%%%%% Appendix B %%%%%
\section{$\vev{W(k)O_{\mu \nu}(k')}$}
\setcounter{equation}{0}
We present some details of the computations
of $\vev{W(k)O_{\mu \nu}(k')}$ given as
(\ref{diagram-1}) through (\ref{diagram-5}) in Subsection 3.3.

\noindent
{\bf Diagram 1}\\
The contribution to $\vev{W(k) O_{\mu \nu}(k')}$ from this diagram
is given by
\begin{eqnarray}
  && \vev{W(k) O_{\mu \nu}(k')}_{\rm Diagram~1}
\nonumber \\
  && = \frac{(ig)^2}{2} \int_\epsilon^{1-\epsilon} d \sigma
    \int \frac{d^3 p}{(2 \pi)^3}~
    e^{-i k \times p \sigma + \frac{i}{2} k \times p}
\nonumber \\ && \qquad \qquad \times
    \vev{l^{\mu_1} A_{\mu_1}(p) l^{\mu_2} A_{\mu_2}(k-p)
         (-i k'_\mu) A_\nu (k')}
    - ( \mu \leftrightarrow \nu ).
\end{eqnarray}
This can be easily evaluated
using (\ref{vertex}) and (\ref{4-epsilons}) as follows:
\begin{eqnarray}
  && \vev{W(k) O_{\mu \nu}(k')}_{\rm Diagram~1}
\nonumber \\
  && = -\frac{(ig)^3}{2} (2 \pi)^3 \delta^{(3)} (k+k')
    \int_\epsilon^{1-\epsilon} d \sigma
    \int \frac{d^3 p}{(2 \pi)^3}~
    \frac{e^{-i k \times p \sigma} - e^{i k \times p (1-\sigma)}}
         {p^2 (k-p)^2}
\nonumber \\ && \qquad \qquad \times
    \left\{ (k \times p) (k_\mu l_\nu -  k_\nu l_\mu)
             - l^2 (k_\mu p_\nu - k_\nu p_\mu) \right\},
\end{eqnarray}
using $l \cdot p = k \times p$.
The part involving $e^{i k \times p (1-\sigma)}$ can be
transformed to the part involving $e^{i k \times p \sigma}$
by the change of variables $\sigma' = 1- \sigma$ and $p' = k-p$.
Therefore, we have
\begin{eqnarray}
  && \vev{W(k) O_{\mu \nu}(k')}_{\rm Diagram~1}
\nonumber \\
  && = -(ig)^3 (2 \pi)^3 \delta^{(3)} (k+k')
    \int_\epsilon^{1-\epsilon} d \sigma
    \int \frac{d^3 p}{(2 \pi)^3}~
    \frac{e^{-i k \times p \sigma}}{p^2 (k-p)^2}
\nonumber \\ && \qquad \qquad \times
    \left\{ (k \times p) (k_\mu l_\nu -  k_\nu l_\mu)
             - l^2 (k_\mu p_\nu - k_\nu p_\mu) \right\}.
\end{eqnarray}

\noindent
{\bf Diagram 2}\\
The contribution to $\vev{W(k) O_{\mu \nu}(k')}$ from this diagram
is given by
\begin{eqnarray}
  && \vev{W(k) O_{\mu \nu}(k')}_{\rm Diagram~2}
\nonumber \\
  && = \frac{(ig)^2}{2} \int_\epsilon^{1-\epsilon} d \sigma
    \int \frac{d^3 p}{(2 \pi)^3}~
    e^{-i k \times p \sigma + \frac{i}{2} k \times p}~
    (-ig) \int \frac{d^3 q}{(2 \pi)^3}~
    e^{-i k' \times q \sigma + \frac{i}{2} k' \times q}
\nonumber \\ && \qquad \qquad \times
    \vev{l^{\mu_1} A_{\mu_1}(p) l^{\mu_2} A_{\mu_2}(k-p)
         A_\mu (q) A_\nu (k'-q)}
    - ( \mu \leftrightarrow \nu ).
\end{eqnarray}
There are two nonvanishing ways
to contract the four gauge fields.
Using the propagator (\ref{propagator}), it is evaluated as
\begin{eqnarray}
  && \vev{W(k) O_{\mu \nu}(k')}_{\rm Diagram~2}
\nonumber \\
  && = -\frac{(ig)^3}{2} (2 \pi)^3 \delta^{(3)} (k+k')
    \int_\epsilon^{1-\epsilon} d \sigma
    \int \frac{d^3 p}{(2 \pi)^3}
    \left\{
    e^{i k \times p (1-\sigma-\epsilon)}
    - e^{-i k \times p (\sigma-\epsilon)}
    \right\}
\nonumber \\ && \qquad \qquad {} \times
    l^{\mu_1} \epsilon_{\mu_1 \rho_1 \mu}
    l^{\mu_2} \epsilon_{\mu_2 \rho_2 \nu}
    \frac{p^{\rho_1} (k-p)^{\rho_2}}{p^2 (k-p)^2}
    - ( \mu \leftrightarrow \nu ).
\end{eqnarray}
Using the identity (\ref{2-epsilons}),
this can be further evaluated as
\begin{eqnarray}
  && \vev{W(k) O_{\mu \nu}(k')}_{\rm Diagram~2}
\nonumber \\
  && = -\frac{(ig)^3}{2} (2 \pi)^3 \delta^{(3)} (k+k')
    \int_\epsilon^{1-\epsilon} d \sigma
    \int \frac{d^3 p}{(2 \pi)^3}
    \frac{e^{i k \times p (1-\sigma-\epsilon)}
          - e^{-i k \times p (\sigma-\epsilon)}}
         {p^2 (k-p)^2}
\nonumber \\ && \qquad \qquad \times
    \left\{
    (k \times p) (k_\mu l_\nu - k_\nu l_\mu)
    - l^2 (k_\mu p_\nu - k_\nu p_\mu) \right\}
\nonumber \\
  && = (ig)^3 (2 \pi)^3 \delta^{(3)} (k+k')
    \int_\epsilon^{1-\epsilon} d \sigma
    \int \frac{d^3 p}{(2 \pi)^3}
    \frac{e^{-i k \times p (\sigma-\epsilon)}}
         {p^2 (k-p)^2}
\nonumber \\ && \qquad \qquad \times
    \left\{
    (k \times p) (k_\mu l_\nu - k_\nu l_\mu)
    - l^2 (k_\mu p_\nu - k_\nu p_\mu) \right\}
\nonumber \\
  && = (ig)^3 (2 \pi)^3 \delta^{(3)} (k+k')
    \int_0^{1-2 \epsilon} d \sigma
    \int \frac{d^3 p}{(2 \pi)^3}
    \frac{e^{-i k \times p \sigma}}
         {p^2 (k-p)^2}
\nonumber \\ && \qquad \qquad \times
    \left\{
    (k \times p) (k_\mu l_\nu - k_\nu l_\mu)
    - l^2 (k_\mu p_\nu - k_\nu p_\mu) \right\},
\end{eqnarray}
where we have used the same change of variables as before.
\\

\noindent
{\bf Diagram 3}\\
The contribution to $\vev{W(k) O_{\mu \nu}(k')}$ from this diagram
is given by
\begin{eqnarray}
  && \vev{W(k) O_{\mu \nu}(k')}_{\rm Diagram~3}
\nonumber \\
  && = (ig)^2 \int_\epsilon^{1-\epsilon} d \sigma
    \int \frac{d^3 p}{(2 \pi)^3}~
    e^{-i k' \times p \sigma + \frac{i}{2} k' \times p}
\nonumber \\ && \qquad \qquad \times
    \vev{l^{\mu_1} A_{\mu_1}(k) l'^{\mu_2} A_{\mu_2}(p)
         (-i)(k'-p)_\mu A_\nu (k'-p)}
    - ( \mu \leftrightarrow \nu ),
\end{eqnarray}
where $l'^\mu = (k' \theta)^\mu = k'_\nu \theta^{\nu \mu}$.
As in the case of Diagram 1, this can be evaluated
using (\ref{vertex}) and (\ref{4-epsilons}) as follows:
\begin{eqnarray}
  && \vev{W(k) O_{\mu \nu}(k')}_{\rm Diagram~3}
\nonumber \\
  && = (ig)^3 (2 \pi)^3 \delta^{(3)} (k+k')
    \int_\epsilon^{1-\epsilon} d \sigma
    \int \frac{d^3 p}{(2 \pi)^3}~
    \left\{ e^{i k \times p \sigma}
            - e^{-i k \times p (1-\sigma)} \right\}
\nonumber \\ && \quad \times
  \Biggl[
  \frac{(k \times p)
          \{ (-k^2 -k \cdot p) (k+p)_\mu l_\nu
          + (k \times p) p_\mu k_\nu \}}
       {k^2 p^2 (k+p)^2}
  + \frac{l^2 k_\mu p_\nu}{k^2 p^2} \Biggr]
\nonumber \\ && \qquad \quad
  {}- ( \mu \leftrightarrow \nu )
\nonumber \\
  && = (ig)^3 (2 \pi)^3 \delta^{(3)} (k+k')
    \int_\epsilon^{1-\epsilon} d \sigma
    \int \frac{d^3 p}{(2 \pi)^3}~ e^{i k \times p \sigma}
\nonumber \\ && \quad \times
  \Biggl[
  \frac{k \times p}{k^2 p^2 (k+p)^2}
  \biggl\{ -(k^2 + k \cdot p) ( k_\mu l_\nu - k_\nu l_\mu)
  - (k^2 + 2 k \cdot p) ( p_\mu l_\nu - p_\nu l_\mu)
\nonumber \\ && \qquad \quad
  {}+ 2 (k \times p) (p_\mu k_\nu  - p_\nu k_\mu) \biggr\}
  + \frac{2 l^2 (k_\mu p_\nu - k_\nu p_\mu)}{k^2 p^2} \Biggr],
\end{eqnarray}
where we have used the appropriate change of variables:
either $\sigma'=1-\sigma$, $p'=-k-p$
or $\sigma'=1-\sigma$, $p'=-p$.
Since
\begin{equation}
  (k \times p)~ e^{i k \times p \sigma}
  = -i \frac{\partial}{\partial \sigma}~
       (e^{i k \times p \sigma}),
\end{equation}
only surface terms of the $\sigma$ integral contribute
for the terms proportional to $k \times p$.
We separate the bulk terms and the surface terms as follows:
\begin{eqnarray}
  && \vev{W(k) O_{\mu \nu}(k')}_{\rm Diagram~3}
\nonumber \\
  && = 2 (ig)^3 (2 \pi)^3 \delta^{(3)} (k+k')
    \int_\epsilon^{1-\epsilon} d \sigma
    \int \frac{d^3 p}{(2 \pi)^3}~ e^{i k \times p \sigma}~
    \frac{l^2 (k_\mu p_\nu - k_\nu p_\mu)}{k^2 p^2}
\nonumber \\ && \qquad
  -i (ig)^3 (2 \pi)^3 \delta^{(3)} (k+k')
    \int \frac{d^3 p}{(2 \pi)^3}~
    \frac{ e^{i k \times p (1-\epsilon)}
           - e^{i k \times p \epsilon} }
         {k^2 p^2 (k+p)^2}
\nonumber \\ && \qquad \quad \times
  \biggl[ -(k^2 + k \cdot p) ( k_\mu l_\nu - k_\nu l_\mu)
  - (k^2 + 2 k \cdot p) ( p_\mu l_\nu - p_\nu l_\mu)
\nonumber \\ && \qquad \qquad \quad
  {}+ 2 (k \times p) (p_\mu k_\nu  - p_\nu k_\mu) \biggr].
\end{eqnarray}

\noindent
{\bf Diagram 4}\\
In order to evaluate this diagram, the following piece
of $O_{\mu \nu}(k)$ at order $g^2$ is necessary:
\begin{equation}
  -(ig)^2 \int d^3 x \int_{2 \epsilon}^{1-\epsilon} d \sigma
  \left[
  l \cdot A(x+l \sigma) \ast A_\mu (x+l \epsilon) \ast A_\nu (x)
  \ast e^{ikx} \right] - ( \mu \leftrightarrow \nu ).
\end{equation}
In momentum space, it is given by
\begin{eqnarray}
  && -(ig)^2 \int_{2 \epsilon}^{1-\epsilon} d \sigma
  \int \frac{d^3 p_1}{(2 \pi)^3} \frac{d^3 p_2}{(2 \pi)^3}
  \frac{d^3 p_3}{(2 \pi)^3} (2 \pi)^3 \delta^{(3)} (k-p_1-p_2-p_3)
\nonumber \\ && \qquad {} \times
  e^{-\frac{i}{2} (p_1 \times p_2 + p_1 \times p_3 + p_2 \times p_3)
     -i k \times p_1 \sigma -i k \times p_2 \epsilon}~
  l \cdot A(p_1) A_\mu (p_2) A_\nu (p_3)
  - ( \mu \leftrightarrow \nu ).
\nonumber \\
\end{eqnarray}
We contract the gauge field coming from the open Wilson line
with one of the gauge fields in the commutator to give
\begin{eqnarray}
  && -(ig)^2 \int_{2 \epsilon}^{1-\epsilon} d \sigma
  \int \frac{d^3 p}{(2 \pi)^3}
  \left\{ e^{-i k \times p (\sigma -\epsilon)}
          - e^{i k \times p (1-\sigma)} \right\}
  \frac{l^\lambda p^\rho}{p^2}
  \left\{ \epsilon_{\lambda \rho \mu} A_\nu (k)
          - \epsilon_{\lambda \rho \nu} A_\mu (k) \right\}
\nonumber \\
  && = -2 (ig)^2 \int_{2 \epsilon}^{1-\epsilon} d \sigma
  \int \frac{d^3 p}{(2 \pi)^3}~
  e^{-i k \times p (\sigma -\epsilon)}~
  \frac{l^\lambda p^\rho}{p^2}
  \left\{ \epsilon_{\lambda \rho \mu} A_\nu (k)
          - \epsilon_{\lambda \rho \nu} A_\mu (k) \right\}.
\end{eqnarray}
The contribution to $\vev{W(k) O_{\mu \nu}(k')}$ from this diagram
is given by contracting the remaining gauge field in the commutator
with the gauge field from the other open Wilson line:
\begin{eqnarray}
  && \vev{W(k) O_{\mu \nu}(k')}_{\rm Diagram~4}
\nonumber \\
  && = -2 (ig)^3 \int_{2 \epsilon}^{1-\epsilon} d \sigma
  \int \frac{d^3 p}{(2 \pi)^3}~
  e^{-i k' \times p (\sigma -\epsilon)}~
  \frac{l'^\lambda p^\rho}{p^2}
\nonumber \\ && \qquad \times
  \left\{ l^{\mu_1} \epsilon_{\lambda \rho \mu}
           \vev{A_{\mu_1} (k) A_\nu (k')}
          - l^{\mu_1} \epsilon_{\lambda \rho \nu}
           \vev{A_{\mu_1} (k) A_\mu (k')} \right\},
\end{eqnarray}
where $l'^\mu = (k' \theta)^\mu = k'_\nu \theta^{\nu \mu}$ as before.
This can be evaluated
using (\ref{propagator}) and (\ref{2-epsilons}):
\begin{eqnarray}
  && \vev{W(k) O_{\mu \nu}(k')}_{\rm Diagram~4}
\nonumber \\
  && = 2 (ig)^3 (2 \pi)^3 \delta^{(3)} (k+k')
  \int_{2 \epsilon}^{1-\epsilon} d \sigma
  \int \frac{d^3 p}{(2 \pi)^3}~
  \frac{e^{i k \times p (\sigma -\epsilon)}}{k^2 p^2}
\nonumber \\ && \qquad \times
  \left\{ (k \times p) (k_\mu l_\nu - k_\nu l_\mu)
          -l^2 (k_\mu p_\nu - k_\nu p_\mu) \right\}
\nonumber \\
  && = -2 (ig)^3 (2 \pi)^3 \delta^{(3)} (k+k')
  \int_{\epsilon}^{1-2 \epsilon} d \sigma
  \int \frac{d^3 p}{(2 \pi)^3}~
  e^{i k \times p \sigma}~
  \frac{l^2 (k_\mu p_\nu - k_\nu p_\mu)}{k^2 p^2}
\nonumber \\ && \quad
  -2i (ig)^3 (2 \pi)^3 \delta^{(3)} (k+k')
  \int \frac{d^3 p}{(2 \pi)^3}~
  \left\{ e^{i k \times p (1 - 2 \epsilon)}
          - e^{i k \times p \epsilon} \right\}
  \frac{k_\mu l_\nu - k_\nu l_\mu}{k^2 p^2},
\end{eqnarray}
where we have separated the bulk contribution
and the surface contribution as before.
\\

\noindent
{\bf Diagram 5}\\
The contribution to $\vev{W(k) O_{\mu \nu}(k')}$ from this diagram
is given by
\begin{eqnarray}
  && \vev{W(k) O_{\mu \nu}(k')}_{\rm Diagram~5}
\nonumber \\
  && = -(ig)^2 \int \frac{d^3 p}{(2 \pi)^3}~
    e^{-i k' \times p \epsilon + \frac{i}{2} k' \times p}
    \vev{l^{\mu_1} A_{\mu_1}(k) A_\mu (p) A_\nu (k'-p)}
    - ( \mu \leftrightarrow \nu ).
\nonumber \\
\end{eqnarray}
Using (\ref{vertex}) and (\ref{4-epsilons}),
this can be evaluated as follows:
\begin{eqnarray}
  && \vev{W(k) O_{\mu \nu}(k')}_{\rm Diagram~5}
\nonumber \\
  && = -i (ig)^3 (2 \pi)^3 \delta^{(3)} (k+k')
    \int \frac{d^3 p}{(2 \pi)^3}~
    \frac{ e^{i k \times p \epsilon}
           - e^{-i k \times p (1-\epsilon)} }
         {k^2 p^2 (k+p)^2}
\nonumber \\ && \qquad \times
   \biggl[
   -2 (k \times p) (k_\mu p_\nu - k_\nu p_\mu)
   -(2 p^2 + k \cdot p) (l_\mu k_\nu - l_\nu k_\mu)
\nonumber \\ && \qquad \qquad
   {}+ (k^2 + 2 k \cdot p) (l_\mu p_\nu - l_\nu p_\mu)
   \biggr].
\end{eqnarray}
It may not be manifest, but the integrand
excluding $e^{i k \times p \epsilon}
        - e^{-i k \times p (1-\epsilon)}$
is invariant under the change of variables $p'=-k-p$.
This is manifest in the denominator. For the numerator,
we can verify the following identity:
\begin{eqnarray}
   && -2 (k \times p) (k_\mu p_\nu - k_\nu p_\mu)
   -(2 p^2 + k \cdot p) (l_\mu k_\nu - l_\nu k_\mu)
   +(k^2 + 2 k \cdot p) (l_\mu p_\nu - l_\nu p_\mu)
\nonumber \\
   && = -2 (k \times p') (k_\mu p'_\nu - k_\nu p'_\mu)
   -(2 p'^2 + k \cdot p') (l_\mu k_\nu - l_\nu k_\mu)
\nonumber \\ && \qquad
   {}+ (k^2 + 2 k \cdot p') (l_\mu p'_\nu - l_\nu p'_\mu).
\end{eqnarray}
Therefore, we have
\begin{eqnarray}
  && \vev{W(k) O_{\mu \nu}(k')}_{\rm Diagram~5}
\nonumber \\
  && = i (ig)^3 (2 \pi)^3 \delta^{(3)} (k+k')
    \int \frac{d^3 p}{(2 \pi)^3}~
    \frac{ e^{i k \times p (1-\epsilon)}
           - e^{i k \times p \epsilon} }
         {k^2 p^2 (k+p)^2}
\nonumber \\ && \qquad \times
   \biggl[
   -2 (k \times p) (k_\mu p_\nu - k_\nu p_\mu)
   -(2 p^2 + k \cdot p) (l_\mu k_\nu - l_\nu k_\mu)
\nonumber \\ && \qquad \qquad
   {}+ (k^2 + 2 k \cdot p) (l_\mu p_\nu - l_\nu p_\mu)
   \biggr].
\end{eqnarray}

%\newpage

%%%%%%%%%% References %%%%%%%%%%

\renewcommand{\baselinestretch}{0.87}

%\bibliography{draft}

\begin{thebibliography}{10}

%\cite{Seiberg:1999vs}
\bibitem{Seiberg:1999vs}
N.~Seiberg and E.~Witten,
``String theory and noncommutative geometry,''
JHEP {\bf 9909}, 032 (1999)
[arXiv:hep-th/9908142].
%%CITATION = HEP-TH 9908142;%%

%\cite{Okawa:2001mv}
\bibitem{Okawa:2001mv}
Y.~Okawa and H.~Ooguri,
``Exact solution to Seiberg-Witten equation
of noncommutative gauge theory,''
Phys.\ Rev.\ D {\bf 64}, 046009 (2001)
[arXiv:hep-th/0104036].
%%CITATION = HEP-TH 0104036;%%

%\cite{Mukhi:2001vx}
\bibitem{Mukhi:2001vx}
S.~Mukhi and N.~V.~Suryanarayana,
``Gauge-invariant couplings of noncommutative branes
to Ramond-Ramond backgrounds,''
JHEP {\bf 0105}, 023 (2001)
[arXiv:hep-th/0104045].
%%CITATION = HEP-TH 0104045;%%

%\cite{Liu:2001pk}
\bibitem{Liu:2001pk}
H.~Liu and J.~Michelson,
``Ramond-Ramond couplings of noncommutative D-branes,''
Phys.\ Lett.\ B {\bf 518}, 143 (2001)
[arXiv:hep-th/0104139].
%%CITATION = HEP-TH 0104139;%%

%\cite{Ishibashi:1999hs}
\bibitem{Ishibashi:1999hs}
N.~Ishibashi, S.~Iso, H.~Kawai and Y.~Kitazawa,
``Wilson loops in noncommutative Yang-Mills,''
Nucl.\ Phys.\ B {\bf 573}, 573 (2000)
[arXiv:hep-th/9910004].
%%CITATION = HEP-TH 9910004;%%

%\cite{Das:2000md}
\bibitem{Das:2000md}
S.~R.~Das and S.~J.~Rey,
``Open Wilson lines in noncommutative gauge theory
and tomography of holographic dual supergravity,''
Nucl.\ Phys.\ B {\bf 590}, 453 (2000)
[arXiv:hep-th/0008042].
%%CITATION = HEP-TH 0008042;%%

%\cite{Gross:2000ba}
\bibitem{Gross:2000ba}
D.~J.~Gross, A.~Hashimoto and N.~Itzhaki,
``Observables of non-commutative gauge theories,''
Adv.\ Theor.\ Math.\ Phys.\  {\bf 4}, 893 (2000)
[arXiv:hep-th/0008075].
%%CITATION = HEP-TH 0008075;%%

%\cite{Okawa:1999cm}
\bibitem{Okawa:1999cm}
Y.~Okawa,
``Derivative corrections to Dirac-Born-Infeld Lagrangian
and non-commutative gauge theory,''
Nucl.\ Phys.\ B {\bf 566}, 348 (2000)
[arXiv:hep-th/9909132].
%%CITATION = HEP-TH 9909132;%%

%\cite{Okawa:2000em}
\bibitem{Okawa:2000em}
Y.~Okawa and S.~Terashima,
``Constraints on effective Lagrangian of D-branes from
non-commutative gauge theory,''
Nucl.\ Phys.\ B {\bf 584}, 329 (2000)
[arXiv:hep-th/0002194].
%%CITATION = HEP-TH 0002194;%%

%\cite{Das:2001xy}
\bibitem{Das:2001xy}
S.~R.~Das, S.~Mukhi and N.~V.~Suryanarayana,
``Derivative corrections from noncommutativity,''
JHEP {\bf 0108}, 039 (2001)
[arXiv:hep-th/0106024].
%%CITATION = HEP-TH 0106024;%%

%\cite{Grandi:2000av}
\bibitem{Grandi:2000av}
N.~Grandi and G.~A.~Silva,
``Chern-Simons action in noncommutative space,''
Phys.\ Lett.\ B {\bf 507}, 345 (2001)
[arXiv:hep-th/0010113].
%%CITATION = HEP-TH 0010113;%%

%\cite{Polychronakos:2002pm}
\bibitem{Polychronakos:2002pm}
A.~P.~Polychronakos,
``Seiberg-Witten map and topology,''
Annals Phys.\  {\bf 301}, 174 (2002)
[arXiv:hep-th/0206013].
%%CITATION = HEP-TH 0206013;%%

%\cite{Bichl:2000bq}
\bibitem{Bichl:2000bq}
A.~A.~Bichl, J.~M.~Grimstrup, V.~Putz and M.~Schweda,
``Perturbative Chern-Simons theory on noncommutative $R^3$,''
JHEP {\bf 0007}, 046 (2000)
[arXiv:hep-th/0004071].
%%CITATION = HEP-TH 0004071;%%

%\cite{Chen:2000ak}
\bibitem{Chen:2000ak}
G.~H.~Chen and Y.~S.~Wu,
``One-loop shift in noncommutative Chern-Simons coupling,''
Nucl.\ Phys.\ B {\bf 593}, 562 (2001)
[arXiv:hep-th/0006114].
%%CITATION = HEP-TH 0006114;%%

%\cite{Susskind:2001fb}
\bibitem{Susskind:2001fb}
L.~Susskind,
``The quantum Hall fluid and non-commutative Chern Simons theory,''
arXiv:hep-th/0101029.
%%CITATION = HEP-TH 0101029;%%

%\cite{Polychronakos:2001mi}
\bibitem{Polychronakos:2001mi}
A.~P.~Polychronakos,
``Quantum Hall states as matrix Chern-Simons theory,''
JHEP {\bf 0104}, 011 (2001)
[arXiv:hep-th/0103013].
%%CITATION = HEP-TH 0103013;%%

%\cite{Hellerman:2001rj}
\bibitem{Hellerman:2001rj}
S.~Hellerman and M.~Van Raamsdonk,
``Quantum Hall physics equals noncommutative field theory,''
JHEP {\bf 0110}, 039 (2001)
[arXiv:hep-th/0103179].
%%CITATION = HEP-TH 0103179;%%

%\cite{Bergman:2001qg}
\bibitem{Bergman:2001qg}
O.~Bergman, J.~H.~Brodie and Y.~Okawa,
``The stringy quantum Hall fluid,''
JHEP {\bf 0111}, 019 (2001)
[arXiv:hep-th/0107178].
%%CITATION = HEP-TH 0107178;%%

%\cite{Hellerman:2001yv}
\bibitem{Hellerman:2001yv}
S.~Hellerman and L.~Susskind,
``Realizing the quantum Hall system in string theory,''
arXiv:hep-th/0107200.
%%CITATION = HEP-TH 0107200;%%

%\cite{Jackiw:2002pn}
\bibitem{Jackiw:2002pn}
R.~Jackiw, S.~Y.~Pi and A.~P.~Polychronakos,
``Noncommuting gauge fields as a Lagrange fluid,''
Annals Phys.\  {\bf 301}, 157 (2002)
[arXiv:hep-th/0206014].
%%CITATION = HEP-TH 0206014;%%

%\cite{Liu:2000mj}
\bibitem{Liu:2000mj}
H.~Liu,
``$\ast$-Trek II: $\ast_n$ operations, open Wilson lines
and the Seiberg-Witten map,''
Nucl.\ Phys.\ B {\bf 614}, 305 (2001)
[arXiv:hep-th/0011125].
%%CITATION = HEP-TH 0011125;%%

%\cite{Okawa:2000sh}
\bibitem{Okawa:2000sh}
Y.~Okawa and H.~Ooguri,
``How noncommutative gauge theories couple to gravity,''
Nucl.\ Phys.\ B {\bf 599}, 55 (2001)
[arXiv:hep-th/0012218].
%%CITATION = HEP-TH 0012218;%%

%\cite{Asakawa:1999cu}
\bibitem{Asakawa:1999cu}
T.~Asakawa and I.~Kishimoto,
``Comments on gauge equivalence in noncommutative geometry,''
JHEP {\bf 9911}, 024 (1999)
[arXiv:hep-th/9909139].
%%CITATION = HEP-TH 9909139;%%

%\cite{Minwalla:1999px}
\bibitem{Minwalla:1999px}
S.~Minwalla, M.~Van Raamsdonk and N.~Seiberg,
``Noncommutative perturbative dynamics,''
JHEP {\bf 0002}, 020 (2000)
[arXiv:hep-th/9912072].
%%CITATION = HEP-TH 9912072;%%

%\cite{VanRaamsdonk:2000rr}
\bibitem{VanRaamsdonk:2000rr}
M.~Van Raamsdonk and N.~Seiberg,
``Comments on noncommutative perturbative dynamics,''
JHEP {\bf 0003}, 035 (2000)
[arXiv:hep-th/0002186].
%%CITATION = HEP-TH 0002186;%%

%\cite{Filk:dm}
\bibitem{Filk:dm}
T.~Filk,
``Divergencies In A Field Theory On Quantum Space,''
Phys.\ Lett.\ B {\bf 376}, 53 (1996).
%%CITATION = PHLTA,B376,53;%%

%\cite{Pisarski:1985yj}
\bibitem{Pisarski:1985yj}
R.~D.~Pisarski and S.~Rao,
``Topologically Massive Chromodynamics
In The Perturbative Regime,''
Phys.\ Rev.\ D {\bf 32}, 2081 (1985).
%%CITATION = PHRVA,D32,2081;%%

%\cite{Alvarez-Gaume:1989wk}
\bibitem{Alvarez-Gaume:1989wk}
L.~Alvarez-Gaume, J.~M.~Labastida and A.~V.~Ramallo,
``A Note On Perturbative Chern-Simons Theory,''
Nucl.\ Phys.\ B {\bf 334}, 103 (1990).
%%CITATION = NUPHA,B334,103;%%

%\cite{Martin:xv}
\bibitem{Martin:xv}
C.~P.~Martin,
``Dimensional Regularization Of Chern-Simons Field Theory,''
Phys.\ Lett.\ B {\bf 241}, 513 (1990).
%%CITATION = PHLTA,B241,513;%%

%\cite{Das:2001kf}
\bibitem{Das:2001kf}
A.~K.~Das and M.~M.~Sheikh-Jabbari,
``Absence of higher order corrections
to noncommutative Chern-Simons coupling,''
JHEP {\bf 0106}, 028 (2001)
[arXiv:hep-th/0103139].
%%CITATION = HEP-TH 0103139;%%

%\cite{Martin:2001ci}
\bibitem{Martin:2001ci}
C.~P.~Martin,
``Computing noncommutative Chern-Simons theory
radiative corrections on the back of an envelope,''
Phys.\ Lett.\ B {\bf 515}, 185 (2001)
[arXiv:hep-th/0104091].
%%CITATION = HEP-TH 0104091;%%

\bibitem{Kaminsky}
K.~Kaminsky, work in progress.
%%CITATION = NONE;%%

%\cite{Sheikh-Jabbari:2001au}
\bibitem{Sheikh-Jabbari:2001au}
M.~M.~Sheikh-Jabbari,
``A note on noncommutative Chern-Simons theories,''
Phys.\ Lett.\ B {\bf 510}, 247 (2001)
[arXiv:hep-th/0102092].
%%CITATION = HEP-TH 0102092;%%

%\cite{Nair:2001rt}
\bibitem{Nair:2001rt}
V.~P.~Nair and A.~P.~Polychronakos,
``On level quantization for the noncommutative
Chern-Simons theory,''
Phys.\ Rev.\ Lett.\  {\bf 87}, 030403 (2001)
[arXiv:hep-th/0102181].
%%CITATION = HEP-TH 0102181;%%

%\cite{Bak:2001ze}
\bibitem{Bak:2001ze}
D.~Bak, K.~M.~Lee and J.~H.~Park,
``Chern-Simons theories on noncommutative plane,''
Phys.\ Rev.\ Lett.\  {\bf 87}, 030402 (2001)
[arXiv:hep-th/0102188].
%%CITATION = HEP-TH 0102188;%%

%\cite{Harvey:2001pd}
\bibitem{Harvey:2001pd}
J.~A.~Harvey,
``Topology of the gauge group in noncommutative gauge theory,''
arXiv:hep-th/0105242.
%%CITATION = HEP-TH 0105242;%%

\end{thebibliography}
%\bibliographystyle{ssg}
\begingroup\raggedright\endgroup
\end{document}